\documentclass[10pt,conference,letterpaper]{IEEEtran}
\usepackage[acronyms,nonumberlist,nopostdot,nomain,nogroupskip,hyperfirst=false]{glossaries}
\AtBeginDocument{%
  \providecommand\BibTeX{{%
    \normalfont B\kern-0.5em{\scshape i\kern-0.25em b}\kern-0.8em\TeX}}}
\usepackage[table,xcdraw]{xcolor}
\usepackage{amssymb}
\usepackage{booktabs}
\usepackage{cite}
\usepackage{graphicx}
\usepackage{epstopdf}
\usepackage{diagbox}
\usepackage{tikz}
\usepackage{arydshln}
\usepackage{pgfplots}
\pgfplotsset{compat=newest}
\pgfplotsset{plot coordinates/math parser=false}
\newlength\fheight
\newlength\fwidth
\usetikzlibrary{plotmarks,patterns,decorations.pathreplacing,backgrounds,calc,arrows,arrows.meta,spy,matrix}
\usepgfplotslibrary{patchplots,groupplots,colorbrewer}
\usepackage{tikzscale}

\newif\ifexttikz
\exttikzfalse

\usepackage{comment}
\usepackage{lipsum}
\usepackage{graphicx}
\usepackage[utf8]{inputenc}
\usepackage{enumitem}
\usepackage{amsmath}
\usepackage{amsfonts}
\usepackage{bm}
\usepackage{tabularx}
\usepackage{svg}
\usepackage{algpseudocode}

\usepackage[linesnumbered,noend]{algorithm2e}

\ifexttikz
	\usetikzlibrary{external}
\fi

\usepackage{multirow}

\renewcommand{\figurename}{Fig.}
\usepackage[font=footnotesize]{subcaption}
\usepackage[font=footnotesize]{caption}

\usepackage{mathtools}

\usepackage{multirow}
\usepackage{url}
\usepackage{tabularx}
\usepackage{array}
\usepackage{multibib}
\newcites{methods}{References}

\newacronym{6g}{6G}{sixth generation}
\newacronym{3gpp}{3GPP}{3rd Generation Partnership Project}
\newacronym{adc}{ADC}{Analog to Digital Converter}
\newacronym{dac}{DAC}{Digital to Analog Converter}
\newacronym{5g}{5G}{5th generation}
\newacronym{aimd}{AIMD}{Additive Increase Multiplicative Decrease}
\newacronym{am}{AM}{Acknowledged Mode}
\newacronym{amc}{AMC}{Adaptive Modulation and Coding}
\newacronym{aoa}{AoA}{Angle of Arrival}
\newacronym{aod}{AoD}{Angle of Departure}
\newacronym{aqm}{AQM}{Active Queue Management}
\newacronym{awgn}{AGWN}{additive white Gaussian noise}
\newacronym{balia}{BALIA}{Balanced Link Adaptation}
\newacronym{bdp}{BDP}{Bandwidth-Delay Product}
\newacronym{fpga}{FPGA}{field-programmable gate array}
\newacronym{cc}{CC}{Congestion Control}
\newacronym{cdf}{CDF}{Cumulative Distribution Function}
\newacronym{cn}{CN}{Core Network}
\newacronym{cm}{CM}{confusion matrix}
\newacronym[plural=\gls{cnn}s,firstplural=convolutional neural networks (CNNs)]{cnn}{CNN}{convolutional neural network}
\newacronym{cqi}{CQI}{Channel Quality Information}
\newacronym{cp}{CP}{Control Plane}
\newacronym{csirs}{CSI-RS}{Channel State Information - Reference Signal}
\newacronym{dc}{DC}{Dual Connectivity}
\newacronym{dce}{DCE}{Direct Code Execution}
\newacronym{dci}{DCI}{Downlink Control Information}
\newacronym{dmr}{DMR}{Deadline Miss Ratio}
\newacronym{dmrs}{DMRS}{DeModulation Reference Signal}
\newacronym{e2e}{E2E}{End-to-End}
\newacronym{ecn}{ECN}{Explicit Congestion Notification}
\newacronym{ebs}{EBS}{exhaustive beam sweep}
\newacronym{edf}{EDF}{Earliest Deadline First}
\newacronym{enb}{eNB}{evolved Node Base}
\newacronym{epc}{EPC}{Evolved Packet Core}
\newacronym{es}{ES}{Edge Server}
\newacronym{fdma}{FDMA}{Frequency Division Multiple Access}
\newacronym{fdd}{FDD}{Frequency Division Duplexing}
\newacronym[firstplural=Radio Access Technologies (RATs)]{rat}{RAT}{Radio Access Technology}
\newacronym{fs}{FS}{Fast Switching}
\newacronym{txer}{TX}{transmitter}
\newacronym{rxer}{RX}{receiver}
\newacronym{bt}{BT}{beam tracking}
\newacronym{ftp}{FTP}{File Transfer Protocol}
\newacronym{gnb}{gNB}{Next Generation Node Base}
\newacronym{bs}{BS}{base station}
\newacronym{harq}{HARQ}{Hybrid Automatic Repeat reQuest}
\newacronym{hetnet}{HetNet}{Heterogeneous Network}
\newacronym{hh}{HH}{Hard Handover}
\newacronym{hol}{HOL}{Head-of-Line}
\newacronym{ia}{IA}{initial access}
\newacronym{imt}{IMT}{International Mobile Telecommunication}
\newacronym{iot}{IoT}{Internet of Things}
\newacronym{los}{LOS}{Line-of-Sight}
\newacronym{lte}{LTE}{Long-Term Evolution}
\newacronym{m2m}{M2M}{Machine to Machine}
\newacronym{ml}{ML}{machine learning}
\newacronym{dl}{DL}{deep learning}
\newacronym{mac}{MAC}{Medium Access Control}
\newacronym{mc}{MC}{Multi-Connectivity}
\newacronym{mcs}{MCS}{Modulation and Coding Scheme}
\newacronym{mec}{MEC}{Mobile Edge Cloud}
\newacronym{mi}{MI}{Mutual Information}
\newacronym{mimo}{MIMO}{multiple-input and multiple-output}
\newacronym{mmwave}{mmWave}{millimeter wave}
\newacronym{mmWave}{mmWave}{Millimeter wave}
\newacronym{mptcp}{MPTCP}{Multipath TCP}
\newacronym{mr}{MR}{Maximum Rate}
\newacronym{mss}{MSS}{Maximum Segment Size}
\newacronym{mtd}{MTD}{Machine-Type Device}
\newacronym{mtu}{MTU}{Maximum Transmission Unit}
\newacronym{nfv}{NFV}{Network Function Virtualization}
\newacronym{nlos}{NLOS}{Non-Line-of-Sight}
\newacronym{nr}{NR}{New Radio}
\newacronym{ofdm}{OFDM}{orthogonal frequency-division multiplexing}
\newacronym{pdcch}{PDCCH}{Physical Downlink Control Channel}
\newacronym{pdcp}{PDCP}{Packet Data Convergence Protocol}
\newacronym{pdsch}{PDSCH}{Physical Downlink Shared Channel}
\newacronym{pdu}{PDU}{Packet Data Unit}
\newacronym{pf}{PF}{Proportional Fair}
\newacronym{pgw}{PGW}{Packet Gateway}
\newacronym{phy}{PHY}{physical layer}
\newacronym{pbch}{PBCH}{Physical Broadcast Channel}
\newacronym[plural=\gls{mme}s,firstplural=Mobility Management Entities (MMEs)]{mme}{MME}{Mobility Management Entity}
\newacronym{prb}{PRB}{Physical Resource Block}
\newacronym{pss}{PSS}{Primary Synchronization Signal}
\newacronym{pucch}{PUCCH}{Physical Uplink Control Channel}
\newacronym{pusch}{PUSCH}{Physical Uplink Shared Channel}
\newacronym{rach}{RACH}{Random Access Channel}
\newacronym{ran}{RAN}{Radio Access Network}
\newacronym{red}{RED}{Random Early Detection}
\newacronym{rf}{RF}{Radio Frequency}
\newacronym{rlc}{RLC}{Radio Link Control}
\newacronym{rlf}{RLF}{Radio Link Failure}
\newacronym{rrc}{RRC}{Radio Resource Control}
\newacronym{rrm}{RRM}{Radio Resource Management}
\newacronym{rr}{RR}{Round Robin}
\newacronym{rs}{RS}{Remote Server}
\newacronym{rsrp}{RSRP}{Reference Signal Received Power}
\newacronym{rss}{RSS}{received signal strength}
\newacronym{rtt}{RTT}{Round Trip Time}
\newacronym{rw}{RW}{Receive Window}
\newacronym{rx}{RX}{Receiver}
\newacronym{sa}{SA}{standalone}
\newacronym{sack}{SACK}{Selective Acknowledgment}
\newacronym{sap}{SAP}{Service Access Point}
\newacronym{ap}{AP}{access point}
\newacronym{sch}{SCH}{Secondary Cell Handover}
\newacronym{scoot}{SCOOT}{Split Cycle Offset Optimization Technique}
\newacronym{sdma}{SDMA}{Spatial Division Multiple Access}
\newacronym{sinr}{SINR}{signal-to-interference-plus-noise ratio}
\newacronym{sm}{SM}{Saturation Mode}
\newacronym{snr}{SNR}{signal-to-noise-ratio}
\newacronym{son}{SON}{Self-Organizing Network}
\newacronym{ss}{SS}{Synchronization Signal}
\newacronym{ssbs}{SSBs}{synchronization signal blocks}
\newacronym{ssb}{SSB}{synchronization signal block}
\newacronym{srs}{SRS}{Sounding Reference Signal}
\newacronym{sss}{SSS}{Secondary Synchronization Signal}
\newacronym{tb}{TB}{Transport Block}
\newacronym{tcp}{TCP}{Transmission Control Protocol}
\newacronym{tdd}{TDD}{Time Division Duplexing}
\newacronym{tdma}{TDMA}{Time Division Multiple Access}
\newacronym{tfl}{TfL}{Transport for London}
\newacronym{tm}{TM}{Transparent Mode}
\newacronym{trp}{TRP}{Transmitter Receiver Pair}
\newacronym{tti}{TTI}{Transmission Time Interval}
\newacronym{ttt}{TTT}{Time-to-Trigger}
\newacronym{tx}{TX}{Transmitter}
\newacronym{ue}{UE}{User Equipment}
\newacronym{ul}{UL}{Uplink}
\newacronym{uml}{UML}{Unified Modeling Language}
\newacronym{um}{UM}{Unacknowledged Mode}
\newacronym{utc}{UTC}{Urban Traffic Control}
\newacronym{vm}{VM}{Virtual Machine}
\newacronym{rsrq}{RSRQ}{Reference Signal Received Quality}
\newacronym{rssi}{RSSI}{Received Signal Strength Indicator}
\newacronym{crs}{CRS}{Cell Reference Signal}
\newacronym{nsa}{NSA}{Non Stand Alone}
\newacronym{mrdc}{MR-DC}{Multi \gls{rat} \gls{dc}}
\newacronym{endc}{EN-DC}{E-UTRAN-\gls{nr} \gls{dc}}
\newacronym{5gc}{5GC}{5G Core}
\newacronym{si}{SI}{Study Item}
\newacronym{iab}{IAB}{Integrated Access and Backhaul}
\newacronym{wf}{WF}{Wired-first}
\newacronym{hqf}{HQF}{Highest-quality-first}
\newacronym{pa}{PA}{Position-aware}
\newacronym{mlr}{MLR}{Maximum-local-rate}
\newacronym{wbf}{WBF}{Wired Bias Function}
\newacronym{mib}{MIB}{Master Information Block}
\newacronym{sib}{SIB}{Secondary Information Block}
\newacronym{kpi}{KPI}{Key Performance Indicator}
\newacronym{ppp}{PPP}{Poisson Point Process}
\newacronym{gtp}{GTP}{GPRS Tunneling Protocol}
\newacronym{amf}{AMF}{Access and Mobility Management Function}
\newacronym{dash}{DASH}{Dynamic Adaptive Streaming over HTTP}
\newacronym{http}{HTTP}{HyperText Transfer Protocol}
\newacronym{qos}{QoS}{quality of service}
\newacronym{udp}{UDP}{User Datagram Protocol}
\newacronym{cu}{CU}{Central Unit}
\newacronym{du}{DU}{Distributed Unit}
\newacronym{mt}{MT}{Mobile Termination}
\newacronym{sdap}{SDAP}{Service Data Adaptation Protocol}
\newacronym{tdm}{TDM}{Time Division Multiplexing}
\newacronym{fdm}{FDM}{Frequency Division Multiplexing}
\newacronym{sdm}{SDM}{Space Division Multiplexing}
\newacronym{dag}{DAG}{Directed Acyclic Graph}
\newacronym{st}{ST}{Spanning Tree}
\newacronym{ummimo}{UM-MIMO}{Ultra-massive Multiple Input, Multiple Output}
\newacronym{uavs}{UAVs}{Unmanned Aerial Vehicles}
\newacronym{wlan}{WLAN}{wireless LAN}
\newacronym{rlnc}{RLNC}{Random Linear Network Coding}
\newacronym{drx}{DRX}{Discontinuous Reception}
\newacronym{cpu}{CPU}{Central Processing Unit}
\newacronym{txb}{TXB}{transmitter's beam}
\newacronym{rxb}{RXB}{receiver's beam}
\newacronym{sifs}{SIFS}{Short Interframe Space}
\newacronym{difs}{DIFS}{DCF Interframe Space}

\newacronym{rfid}{RFID}{Radio Frequency Identification}
\newacronym{rfp}{RFP}{radio fingerprinting}
\newacronym{sdr}{SDR}{software-defined radio}
\newacronym{dnn}{DNN}{deep neural network}

\newacronym{od}{OD}{object detection}

\newacronym{ot}{OT}{object tracking}

\newacronym{har}{HAR}{human activity recognition}
\newacronym{csi}{CSI}{channel state information}
\newacronym{cfr}{CFR}{channel frequency response}
\newacronym{fsl}{FSL}{few-shot learning}
\newacronym{pwr}{PWR}{passive Wi-Fi radar}
\newacronym{matnet}{MatNet}{matching network}
\newacronym{prnet}{ProtoNet}{prototypical network}
\newacronym{lstm}{LSTM}{long short-term memory}
\newacronym{gan}{GAN}{generative adversarial network}
\newacronym{svd}{SVD}{singular value decomposition}
\newacronym{nic}{NIC}{network interface card}
\newacronym{sgd}{SGD}{stochastic gradient descend }
\newacronym{vht}{VHT}{very-high-throughput}
\newacronym{sc}{SC}{split computing}
\newacronym{mumimo}{MU-MIMO}{multi-user multiple-input and multiple-output}
\newacronym{cs}{CS}{compressive sensing}
\newacronym{gr}{GR}{Givens rotation}
\newacronym{ae}{AE}{autoencoder}
\newacronym{ber}{BER}{bit error rate}
\newacronym{sta}{STA}{station}
\newacronym{ndp}{NDP}{null data packet}
\newacronym{bm}{BM}{beamforming matrix}
\newacronym{iui}{IUI}{inter-user interference}
\newacronym{bmr}{BMR}{\gls{bm} report}
\newacronym{sumimo}{SU-MIMO}{single-user MIMO}
\newacronym{flops}{FLOPS}{floating point operations per second}
\newacronym{mse}{MSE}{mean-squared error}
\newacronym{dsp}{DSP}{digital signal processing}
\newacronym{bf}{BF}{beamforming feedback}
\tikzstyle{startstop} = [rectangle, rounded corners, minimum width=2cm, minimum height=0.5cm,text centered, draw=black]
\tikzstyle{io} = [trapezium, trapezium left angle=70, trapezium right angle=110, minimum width=3cm, minimum height=1cm, text centered, draw=black]
\tikzstyle{process} = [rectangle, minimum width=2cm, minimum height=0.5cm, text centered, draw=black, alignb=center]
\tikzstyle{decision} = [ellipse, minimum width=2cm, minimum height=1cm, text centered, draw=black]
\tikzstyle{arrow} = [thick,<->,>=stealth]
\tikzstyle{line} = [thick,>=stealth]
\tikzstyle{darrow} = [thick,<->,>=stealth,dashed]
\tikzstyle{sarrow} = [thick,->,>=stealth]
\tikzstyle{larrow} = [line width=0.1mm,dashdotted,->,>=stealth]

\makeatletter
\def\grd@save@target#1{%
  \def\grd@target{#1}}
\def\grd@save@start#1{%
  \def\grd@start{#1}}
\tikzset{
  grid with coordinates/.style={
    to path={%
      \pgfextra{%
        \edef\grd@@target{(\tikztotarget)}%
        \tikz@scan@one@point\grd@save@target\grd@@target\relax
        \edef\grd@@start{(\tikztostart)}%
        \tikz@scan@one@point\grd@save@start\grd@@start\relax
        \draw[minor help lines] (\tikztostart) grid (\tikztotarget);
        \draw[major help lines] (\tikztostart) grid (\tikztotarget);
        \grd@start
        \pgfmathsetmacro{\grd@xa}{\the\pgf@x/1cm}
        \pgfmathsetmacro{\grd@ya}{\the\pgf@y/1cm}
        \grd@target
        \pgfmathsetmacro{\grd@xb}{\the\pgf@x/1cm}
        \pgfmathsetmacro{\grd@yb}{\the\pgf@y/1cm}
        \pgfmathsetmacro{\grd@xc}{\grd@xa + \pgfkeysvalueof{/tikz/grid with coordinates/major step x}}
        \pgfmathsetmacro{\grd@yc}{\grd@ya + \pgfkeysvalueof{/tikz/grid with coordinates/major step y}}
        \foreach \x in {\grd@xa,\grd@xc,...,\grd@xb}
        \node[anchor=north] at (\x,\grd@ya) {\pgfmathprintnumber{\x}};
        \foreach \y in {\grd@ya,\grd@yc,...,\grd@yb}
        \node[anchor=east] at (\grd@xa,\y) {\pgfmathprintnumber{\y}};
      }
    }
  },
  minor help lines/.style={
    help lines,
    gray,
    line cap =round,
    xstep=\pgfkeysvalueof{/tikz/grid with coordinates/minor step x},
    ystep=\pgfkeysvalueof{/tikz/grid with coordinates/minor step y}
  },
  major help lines/.style={
    help lines,
    line cap =round,
    line width=\pgfkeysvalueof{/tikz/grid with coordinates/major line width},
    xstep=\pgfkeysvalueof{/tikz/grid with coordinates/major step x},
    ystep=\pgfkeysvalueof{/tikz/grid with coordinates/major step y}
  },
  grid with coordinates/.cd,
  minor step x/.initial=.5,
  minor step y/.initial=.2,
  major step x/.initial=1,
  major step y/.initial=1,
  major line width/.initial=1pt,
}
\makeatother

\usepackage{comment}
\usepackage{xfrac}
\usepackage{xspace}

\begin{document}

\newcommand{\NW}{\texttt{SplitBeam}\xspace}

\title{\NW: Effective and Efficient Beamforming in Wi-Fi Networks Through Split Computing\vspace{-0.3cm}}


\author{\IEEEauthorblockN{Niloofar Bahadori$^*$, Yoshitomo Matsubara$^{\dag}$, Marco Levorato$^{\dag}$, and Francesco Restuccia$^{*}$\vspace{-0.4cm}}\\
 \IEEEauthorblockA{$^\dag$ Donald Bren School of Information and Computer Sciences, University of California at Irvine, United States}
 \IEEEauthorblockA{$^{*}$ Institute for the Wireless Internet of Things, Northeastern University, United States}
 \IEEEauthorblockA{Corresponding author e-mail: \texttt{frestuc@northeastern.edu}}
 \vspace{-0.9cm}}

\maketitle

\glsdisablehyper
\begin{abstract}
Modern IEEE 802.11 (Wi-Fi) networks extensively rely on multiple-input multiple-output (MIMO) to significantly improve throughput. To correctly beamform MIMO transmissions, the access point  needs to frequently acquire a beamforming matrix (BM) from each connected station. However, the size of the matrix grows with the number of antennas and subcarriers, resulting in an increasing amount of airtime overhead and computational load at the station. Conventional approaches come with either excessive computational load or loss of beamforming precision. For this reason, we propose SplitBeam, a new framework where we train a split deep neural network (DNN) to directly output the BM given the channel state information (CSI) matrix as input. The DNN is designed with an additional ``bottleneck'' layer to ``split'' the original DNN into a head model and a tail model, respectively executed by the station and the access point. The head model generates a compressed representation of the BM, which is then used by the AP to produce the BM using the tail model. We formulate and solve a bottleneck optimization problem (BOP) to keep computation, airtime overhead, and bit error rate (BER) below application requirements. We perform extensive experimental CSI collection with off-the-shelf Wi-Fi devices in two distinct environments and compare the performance of SplitBeam with the standard IEEE 802.11 algorithm for BM feedback and the state-of-the-art DNN-based approach LB-SciFi. Our experimental results show that SplitBeam reduces the beamforming feedback size and computational complexity by respectively up to 81\% and 84\% while maintaining BER within about $10^{-3}$ of existing approaches. We also implement the SplitBeam DNNs on FPGA hardware to estimate the end-to-end BM reporting delay, and show that the latter is less than 10 milliseconds in the most complex scenario, which is the target channel sounding frequency in realistic multi-user MIMO scenarios.  To allow full reproducibility, we will release our code and datasets to the community.

\end{abstract}

\maketitle
\pagestyle{plain}

\newcommand\blfootnote[1]{%
  \begingroup
  \renewcommand\thefootnote{}\footnote{#1}%
  \addtocounter{footnote}{-1}%
  \endgroup
}


\glsresetall
\glsdisablehyper


\section{Introduction}

Today, Wi-Fi networks are used to connect hundreds of millions of people worldwide. Wi-Fi is so ubiquitous that cellular operators are expected to offload 63\% of their traffic to Wi-Fi by 2022 \cite{CiscoVNI}. To attest to the need for higher data rates, the IEEE is currently standardizing 802.11be (Wi-Fi 7), which will support throughput of up to 46 Gbps  through wider signal bandwidths and the usage of \gls{mumimo} techniques \cite{deng2020ieee}. MU-MIMO will become fundamental also to effectively decongest the unlicensed spectrum bands through spatial reuse, which are increasingly saturated \cite{SpectrumCrunch}. To correctly beamform transmissions, MU-MIMO requires \glspl{ap} to periodically collect \gls{csi} from each connected \gls{sta} to beamform the transmissions \cite{802.11ac,802.11ax}. According to the IEEE 802.11 standard \cite{perahia2013next}, the \gls{bf} is constructed by (i) measuring the \gls{csi} through pilot signals and  (ii) computing the \gls{bf} through \gls{svd}. Then, the \gls{bf} is decomposed into \gls{gr} angles that produce the \gls{bm}, as explained in Section \ref{sec:prelim}. 

A key challenge in MIMO systems is that the size of the \gls{bf} grows with the number of subcarriers, transmitting and receiving antennas. For example, in an $8 \times 8$ network at 160 MHz of bandwidth, the \gls{bf} in 802.11 will be of size (486 subcarriers $\times$ 56 angles/subcarrier $\times$ 16 bits/angle =)  435,456 bits $\simeq$ 54.43 kB, if the maximum angle resolution is used. If \glspl{bf} are sent back every 10 ms as suggested in \cite{gast2013802}, the airtime overhead is 435,456 / 0.01 $\simeq$  43.55 Mbit/s. Moreover, the \gls{bf} computation imposes a significant burden on the \glspl{sta}, which may become intolerable for low-power devices. Specifically, the complexity of \gls{svd} and \gls{gr} are $\mathcal{O}( (4N_t N_r^2 + 22  N_t^3) \cdot S )$ and $\mathcal{O}( N_t^3 N_r^3 S)$, where $N_t$, $N_r$ and $S$ denote the number of transmitting and receiving antennas and subcarriers \cite{golub1996matrix}. Since Wi-Fi 7 will support more spatial streams (up to 16) and bandwidth (up to 320 MHz), a thorough revision of how MIMO is performed in Wi-Fi is quintessential to keep the complexity under control.    \vspace{-0.2cm}


\begin{figure}[!ht]
    \centering
    \includegraphics[width=0.95\columnwidth]{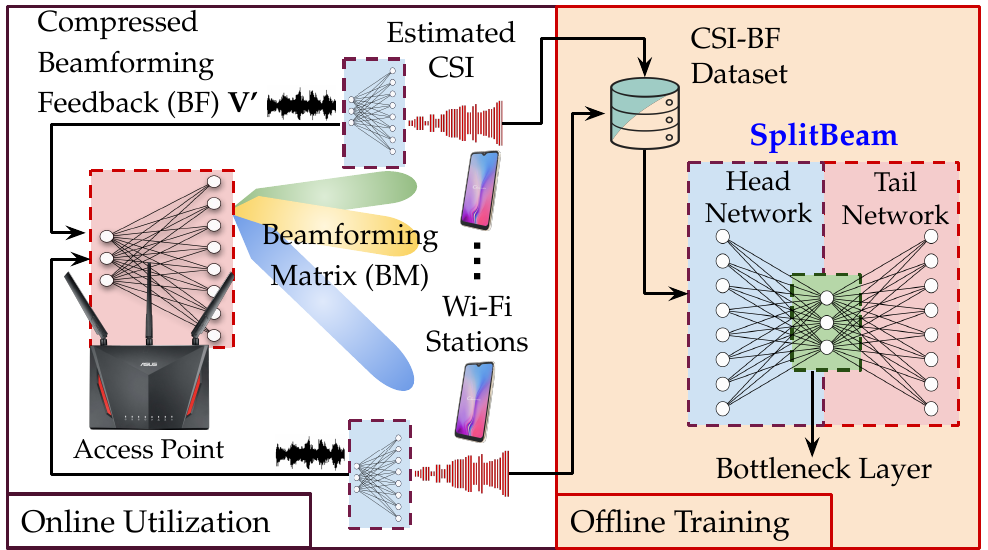}
    \vspace{-0.35em}
    \caption{High-level overview of the operations in \NW.\vspace{-0.2cm}}
    \label{fig:intro}
\end{figure}

Existing approaches to reduce MIMO complexity -- discussed in details in Section \ref{sec:RelatedWorks} -- come with excessive computation overhead and/or performance loss, with most of them not being compliant to the IEEE 802.11 standard \cite{hassan2020performance, statBF2016three,statBFzhang2017sum, monteiro2021massive, metzler2016denoising,azizipour2019compressed,feedbackReduction,lu2018mimo,CSRenet,wen2018deep,mashhadi2020distributed,sangdeh2020lb}. In this paper, we take a different approach and present \NW, an \textit{IEEE 802.11 standard-compliant} framework leveraging split computing to drastically decrease both computational load and \gls{bf} size while maintaining reasonable beamforming accuracy. Figure \ref{fig:intro} shows a high-level overview of \NW. We first train a \gls{dnn} model to map the estimated \gls{csi} matrix to the \gls{bf} in a supervised manner. Second, we ``split'' the \gls{dnn} into a \textit{head} and a \textit{tail} model, respectively executed by the \glspl{sta} and by the \gls{ap}. The head model is custom-trained to produce a compressed representation of the \gls{bf} through the introduction of a ``bottleneck'' inside the model, thus reducing BF airtime and \gls{sta} computational load. 

The key advantage of our approach is that the complexity of the head model and the \gls{bf} representation size can be adjusted by modifying the bottleneck placement and size. Indeed, the bottleneck can trade off computational load, feedback size and beamforming accuracy, which was not available in previous approaches. This  is crucial for constrained Wi-Fi devices and systems, which will cater to heterogeneous devices with different processing capacities \cite{deng2020ieee}.\smallskip

\textbf{This paper makes the following novel contributions:}\smallskip

$\bullet$ We propose \NW, a novel framework for \gls{bf} compression and STA computation reduction in \gls{mumimo} Wi-Fi networks. We perform a complexity analysis in Section \ref{sec:complexity} and show that on average, \NW successfully reduces the STA computational load and the \gls{bf} size by respectively 92\% and 91\% when compared to the standardized 802.11 algorithm;\smallskip   

$\bullet$ We formulate a bottleneck optimization problem (BOP) to determine the bottleneck placement and size with the goal of minimizing airtime and computation overhead, while ensuring that the \gls{ber} and end-to-end delay are below the application's desired level (Section \ref{sec:optimization}). Given its complexity, we introduce a heuristic algorithm and propose a customized training procedure for the resulting \gls{dnn}; \smallskip

$\bullet$ We leverage off-the-shelf Wi-Fi equipment to collect CSI data in two different propagation environments, and compare the performance of \NW with IEEE 802.11ac/ax CSI feedback algorithm \cite{802.11ac,802.11ax} (henceforth called 802.11 for brevity) and the state-of-the-art \gls{dnn}-based compression technique, LB-SciFi \cite{sangdeh2020lb}. Experimental results in Section \ref{sec:results} show that the computational load and feedback size are reduced by up to 84\% and 81\% with respect to 802.11. Also, with the same compression rate, the computational load is reduced by up to 89\% compared to LB-SciFi; \smallskip

$\bullet$ We have synthesized \NW in \gls{fpga} hardware by using a customized library to show the feasibility of \NW in real-world Wi-Fi systems. Our experimental results show that the maximum end-to-end latency incurred by \NW is less than 7 milliseconds (ms) in the case of $4 \times 4$ MIMO operating at 160 MHz and lowest compression rate, which is well below the suggested threshold of 10ms in MU-MIMO Wi-Fi systems \cite{gast2013802}. \textbf{We pledge to release our code and our 230 GB dataset to the community for full reproducibility.}

\section{Background and Related Work} \label{sec:RelatedWorks}

In this section, we discuss prior work and highlight the novelty of this paper. We summarize \gls{csi} compression methods and data-driven feedback techniques in Section 2.1 and 2.2.
\gls{csi} collection methodologies and current \gls{mumimo} \gls{csi} datasets are discussed in Section 2.3. \smallskip

\noindent\textbf{2.1: Traditional CSI Feedback Compression.}~Existing approaches can be categorized into (i) statistical approaches, (ii) \gls{cs}, and (iii) polar decomposition (PD) methods. The former methods leverage channel statistics to reduce the reporting frequency  \cite{hassan2020performance, statBF2016three,statBFzhang2017sum, monteiro2021massive}. As a consequence, their performance deteriorates in dynamic channel environments. Conversely, \gls{cs} takes advantage of the sparsity of the channel response to compress the \gls{csi}. However, indoor channels might not be as sparse due to the presence of multiple reflectors. Moreover, many widely-used \gls{cs} techniques such as BM3D-AMP \cite{metzler2016denoising} and OMP-US \cite{azizipour2019compressed} experience slow convergence time. PD leverages the fact that the \gls{bf} matrix is \textit{unitary}. Thus, approaches such as Givens rotations (GR) can reduce the feedback size. However, computing and compressing the \gls{bf} imposes an additional computational load. Subcarrier grouping, wide-band precoding \cite{feedbackReduction} and reducing the number of feedback bits \cite{perahia2013next} can be used to decrease complexity, which come at the detriment of beamforming accuracy.\smallskip

\noindent\textbf{2.2: Data-Driven CSI Compression.}
\label{sec:csi_learning}
Deep learning (DL) has been used for \gls{sumimo} \gls{csi} compression \cite{lu2018mimo,CSRenet,wen2018deep,mashhadi2020distributed}. For instance, CS-ReNet \cite{CSRenet}, CsiNet\cite{wen2018deep}, DeepCMC \cite{mashhadi2020distributed} have used \gls{cnn} and \gls{lstm} to extract the location of the significant time-domain channel taps by exploiting channel redundancy. However, the above work mostly assumes low-mobility, single-user, and outdoor \gls{lte} scenarios, where channels responses are highly redundant and sparse due to the limited mobility and few local scatters at the \gls{bs}. In addition, conversely from \gls{sumimo}, inaccuracy in the beamforming will lead to \gls{iui} in \gls{mumimo}, which reduces the \gls{sinr} significantly. Therefore, the \gls{csi} must be of higher resolution and more frequently updated. LB-SciFi \cite{sangdeh2020lb} is the first \gls{dl}-based work that  investigated \gls{bf} compression of an indoor \gls{wlan} network through adopting a \gls{ae}-based \gls{dnn}. LB-SciFi model is composed of an encoder and a decoder. This model requires \glspl{sta} to (i) compute the \gls{bf} through \gls{svd}, (ii) decompose the \gls{bf} into $\psi$ and $\phi$ angles using \gls{gr}, and (iii) compress the angles using the encoder. At the \gls{ap}, the received codes are decompressed using the decoder, and further inverse \gls{gr} must be applied to recover the \gls{bf}. Thus, the LB-SciFi's encoder compounds the complexity of \gls{svd} and \gls{gr} operations, which may exclude resource constraint devices. In addition, since the \gls{ae} is trained for 20 MHz channels with 56 subcarriers, the growth rate of the encoder's complexity with respect to the number of subcarriers is unknown. Conversely, in this work, we focus on reducing the computational load, as well as feedback airtime while maintaining the \gls{ber} at an acceptable level. In Section \ref{sec:results}, we show that while \NW achieves the same level of feedback compression as LB-SciFi, it reduces the \glspl{sta}' computational load up to 89\%. \smallskip


\noindent\textbf{2.3: CSI Collection and Datasets Availability.}~To the best of our knowledge, only a few Wi-Fi \gls{csi} datasets are publicly available, with the majority based on simulation data \cite{CSRenet,wen2018deep,mashhadi2020distributed}. Authors in \cite{sangdeh2020lb} collected \gls{mumimo} \gls{csi} dataset for training and evaluating the LB-SciFi. However, the experiment is limited to 20 MHz and the dataset is not publicly available.
Although most commercial Wi-Fi chipsets can potentially generate CSI data, few manufacturers make this data available to developers and researchers, especially for modern chipsets such as 802.11ac/ax. Recently, the Nexmon firmware patch has been released, allowing the extraction of CSI from specific Broadcom/Cypress Wi-Fi chipsets \cite{nexmonCSI}.
To address the lack of large-scale \gls{mumimo} wireless dataset,  for the first time, \textit{we collect a large-scale dataset containing multi-user (up to 3) multi-antenna (up to 3) fine-grained (up to 242 subcarriers) \gls{csi} data from multiple environments with different propagation characteristics using off-the-shelf Wi-Fi routers, which we will release to the community.} \vspace{-0.2cm}

\section{Problem Statement and Challenges} \label{sec:problem_statement}
In this section, we detail the \gls{bf} acquisition procedure in \gls{wlan} 802.11 systems. Then, we discuss the challenges of applying such a technique to next-generation Wi-Fi systems. \vspace{-0.1cm}

\subsection{System Model and Preliminaries} \label{sec:prelim}
In this section, we briefly introduce some terminology. We will adopt the following notation for mathematical expressions. We use boldface uppercase letters to denote matrices. We use the superscripts $T$ and $\dag$ to denote the transpose and the complex conjugate transpose (i.e., the Hermitian). We define with $\angle{\mathbf{C}}$ the matrix containing the phases of the complex-valued matrix $\mathbf{C}$. Moreover, {\rm diag}$(c_1, \dots, c_j)$ indicates the diagonal matrix with elements $(c_1, \dots, c_j)$ on the main diagonal. The $(c_1, c_2)$ entry of matrix $\mathbf{C}$ is defined by $\left[\mathbf{\mathbf{C}}\right]_{c_1, c_2}$, while $\mathbb{I}_{c}$ refers to an identity matrix of size $c \times c$ and $\mathbb{I}_{c\times d}$ is a $c \times d$ generalized identity matrix. The notations $\mathbb{R}$ and $\mathbb{C}$ will indicate the set of real and complex numbers, respectively. 

\subsubsection{WLAN MU-MIMO System Model}
\begin{figure}[!ht]
    \centering
    \includegraphics[width=0.95\columnwidth]{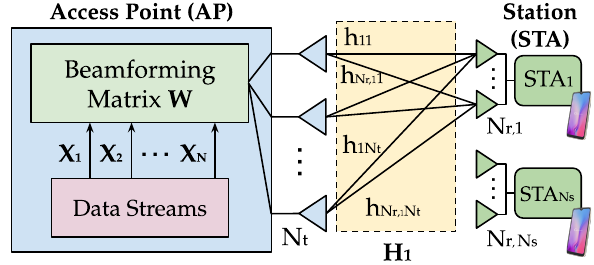}
    \caption{Overview of a Wi-Fi MU-MIMO System.\vspace{-0.3cm}}
    \label{fig:mu-mimo}
\end{figure}
We consider a \gls{mumimo} system with an \gls{ap} as the beamformer, and a set $\mathcal{I}$ of $N_s$ \gls{sta} devices as beamformees. The configuration of the \gls{mumimo} system is shown in Figure \ref{fig:mu-mimo}, where $N_{t}$ antennas are located at the \gls{ap} and $N_{r,i}$ antennas are at each client. $N_{ss,i}$ is the number of spatial streams for \gls{sta} $i$. Let $\mathbf{X}_i(s) \in \mathbb{C}^{N_{ss,i} \times 1}$ represent the transmitted data symbol vector for user $i$ over subcarrier $s \in \mathcal{S}$, where $\mathcal{S}$ is the set of $S$ \gls{ofdm} subcarriers. Each data symbol vector is beamformed through a beamforming matrix (BM) denoted by $\mathbf{W}_i(s)\in \mathbb{C}^{N_t \times  N_{ss, i}}$. By defining the fading channel from the \gls{ap} to \gls{sta} $i$ as $\mathbf{H}_i(s) \in \mathbb{C}^{N_{r, i} \times N_t}$, the received signal at \gls{sta} $i$ is
\begin{equation}
    \mathbf{Y}_i = \sqrt{\frac{\rho}{N_t}} \hspace{1mm} \left( \mathbf{H}_i \mathbf{W}_i \mathbf{X}_i + \sum\limits_{\substack{j\in \mathcal{I}}\setminus i} \mathbf{H}_i \mathbf{W}_j \mathbf{X}_j\right) + \mathbf{N}_i, \label{eq:recSig}
\end{equation}
where $\rho$ denotes the \gls{snr} and is assumed equal for all users. $\mathbf{N}_i$ is the complex \gls{awgn} for \gls{sta} $i$ as $\mathcal{CN}(0,1)$. 
To simplify notation, \eqref{eq:recSig} is given in terms of the frequency domain for a single subcarrier and subcarrier index $(s)$ is omitted.  We assume the number of transmit antennas is set to be the sum total of all the used spatial streams, $N_t = \sum_{i \in \mathcal{I}}N_{ss,i}$. The first term in \eqref{eq:recSig} denotes the desired signal and the second term is the inter-user interference, which can be eliminated thanks to the beamforming. Ideally, $\mathbf{H}_i \mathbf{W}_j = 0$ when $i \neq j$. Therefore, the received signal can be reduced to $\mathbf{Y}_i = \sqrt{\sfrac{\rho}{N_t}} \mathbf{H}_i \mathbf{W}_i \mathbf{X}_i$.


\subsubsection{Computing the Beamforming Matrix} \label{sec:prelim_beamforming}
In \gls{mumimo} Wi-Fi systems, the beamforming matrix $\mathbf{W}$ with dimension $N_t \times \sum_{i=1}^{N_s} N_{ss, i} \times S$ is calculated using a multi-user channel sounding mechanism, shown in Figure \ref{fig:sounding}. The procedure contains three main steps: 
\begin{figure}[!ht]
    \centering
    \includegraphics[width=\columnwidth]{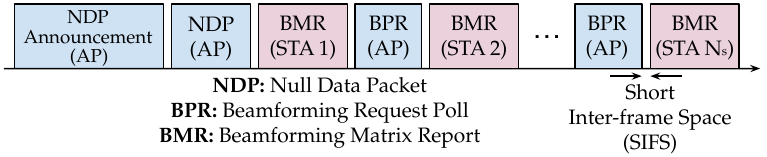}
    \caption{Multi-user channel sounding procedure in IEEE 802.11.\vspace{-0.3cm}}
    \label{fig:sounding}
\end{figure}

\textbf{(1)} The AP begins the process by transmitting a \gls{ndp} announcement frame, used to gain control of the channel and identify the STAs. The AP follows the \gls{ndp} announcement frame with a \gls{ndp} for each spatial stream;\smallskip

\textbf{(2)} Upon reception of the \gls{ndp}, each \gls{sta} $i$ analyzes the \gls{ndp} training fields -- for example, VHT-LTF (Very High Throughput Legacy Training Field) in 802.11ac -- and estimates the channel matrix $\mathbf{H}_i(s)$ for all subcarriers $s$, which is then decomposed by using \gls{svd}:

\begin{equation}
    \mathbf{H}_i(s) = \mathbf{U}_i(s)\cdot \mathbf{S}_i(s) \cdot \mathbf{Z}_i(s)^{\dagger}
\end{equation}
where $\mathbf{U}_i(s) \in \mathbb{C}^{N_{r,i} \times N_{r,i}}$ and $\mathbf{Z}_i(s) \in \mathbb{C}^{N_{t} \times N_{t}}$ are unitary matrices, while the singular values are collected in the $N_{r,i} \times N_{t}$ diagonal matrix $\mathbf{S}_i(s)$. With this notation, the complex-valued BM $\mathbf{V}_i(s)$ is defined by collecting the first $N_{ss,i}$ columns of $\mathbf{Z}_i(s)$. To simplify the notation, we will now drop the $i$ subscript and refer to a generic receiver. To reduce the channel overhead, $\mathbf{V}(s)$ is converted into polar coordinates as detailed in Algorithm~\ref{alg:beamf_feedback}. The output is matrices $\mathbf{D}_{s,t}$ and $\mathbf{G}_{s,\ell,t}$, defined as
\begin{equation}
    \mathbf{D}_{s,t} =
	 \begin{bmatrix}
	\mathbb{I}_{t-1} & 0 & \multicolumn{2}{c}{\dots} & 0 \\
	0 & e^{j\phi_{s,t,t}} & 0 & \dots & \multirow{2}{*}{\vdots} \\
	\multirow{2}{*}{\vdots} & 0 & \ddots & 0 &  \\
	 & \vdots & 0 & e^{j\phi_{s,N_t-1,t}} & 0 \\
	0 & \multicolumn{2}{c}{\dots} & 0 & 1
	 \end{bmatrix},\label{eq:d_matrix}
\end{equation}
\begin{equation}
    \mathbf{G}_{s,\ell,t} =
	 \begin{bmatrix}
	\mathbb{I}_{t-1} & 0 & \multicolumn{2}{c}{\dots} & 0 \\
	0 & \cos{\psi_{s,\ell,t}} & 0 & \sin{\psi_{s,\ell,t}} & \multirow{2}{*}{\vdots} \\
	\multirow{2}{*}{\vdots} & 0 & \mathbb{I}_{\ell-t-1} & 0 &  \\
	 & -\sin{\psi_{s,\ell,t}} & 0 & \cos{\psi_{s,\ell,t}} & 0 \\
	0 & \multicolumn{2}{c}{\dots} & 0 & \mathbb{I}_{N_{r}-\ell}
	 \end{bmatrix},\label{eq:g_matrix}
\end{equation}
that allow rewriting $\mathbf{V}(s)$ as $\mathbf{V}(s) = \mathbf{\Tilde{V}}(s) \cdot \mathbf{\Tilde{D}}(s)$, with
\begin{equation}
    \mathbf{\Tilde{V}}(s) = \prod_{t=1}^{\min(N_{ss}, N_{t}-1)} \Bigg( \mathbf{D}_{s,t} \prod_{l=t+1}^{N_t}\mathbf{G}_{s,l,t}^T\Bigg) \cdot \mathbb{I}_{N_{t} \times N_{ss}}, \label{eq:v_matrix}
\end{equation}
In the $\mathbf{\Tilde{V}}(s)$ matrix, the last row -- i.e., the feedback for the $N_t$-th transmitting antenna -- consists of non-negative real numbers by construction. Using this transformation, the STA is only required to transmit the $\phi$ and $\psi$ angles to the AP. Moreover, it has been proved (see \cite{perahia2013next}, Chapter~13) that the beamforming performance is equivalent when using $\mathbf{V}(s)$ or $\mathbf{\Tilde{V}}(s)$. Thus, the feedback for $\mathbf{\Tilde{D}}_k$ is not fed back to the AP.

\RestyleAlgo{ruled}
\SetKwComment{Comment}{/*}{*/}
\SetAlgoNoLine
\LinesNotNumbered
\begin{algorithm}[h!]
\caption{$\mathbf{V}(s)$ decomposition}\label{alg:beamf_feedback}
Require: $\mathbf{V}(s)$\;
$\mathbf{\Tilde{D}}(s) = {\rm diag}(e^{j \angle \left[\mathbf{V}(s)\right]_{N_t,1}}, \dots, e^{j \angle \left[\mathbf{V}(s)\right]_{N_t,N_{ss}}})$ \;
$\mathbf{\Omega}(s) = \mathbf{V}(s) \cdot \mathbf{\Tilde{D}}(s)^\dag$\;
\For{$t \leftarrow 1$ to $\min (N_{ss}, N_t-1)$}{
$\phi_{s,\ell,t} = \angle \left[\mathbf{\Omega}(s)\right]_{\ell, t}$ with $\ell={t, \dots, N_t-1}$\;
compute $\mathbf{D}_{s,t}$ through Equation~(\ref{eq:d_matrix})\;
$\mathbf{\Omega}(s) \leftarrow \mathbf{D}_{s,t}^\dag \cdot  \mathbf{\Omega}(s)$\;
\For{$\ell \leftarrow t+1$ to $N_t$}{
$\psi_{s,\ell,t} = \arccos \left( \frac{[\mathbf{\Omega}(s)]_{t, t}}{\sqrt{[\mathbf{\Omega}(s)]_{t, t}^2 + [\mathbf{\Omega}(s)]_{\ell, t}^2}} \right)$\;
compute $\mathbf{G}_{s,\ell,t}$ through Equation~(\ref{eq:g_matrix})\;
$\mathbf{\Omega}(s) \leftarrow \mathbf{G}_{s,\ell,t} \cdot \mathbf{\Omega}(s)$\;}}
\end{algorithm}

\smallskip

\textbf{(3)} The AP transmits a beamforming report poll (BRP) frame to retrieve the angles from each STA. The angles are further quantized using $b_{\phi} \in \{7, 9\}$ bits for $\phi$ and $b_{\psi} = b_{\phi}-2$ bits for $\psi$, to further reduce the channel occupancy. The quantized values -- \mbox{$q_{\phi} = \{0, \dots, 2^{b_{\phi}}-1\}$} and \mbox{$q_{\psi} = \{0, \dots, 2^{b_{\psi}}-1\}$} -- are packed into a compressed beamforming frame (CBF). Each  contains $A$ number of angles for each of the $S$ \gls{ofdm} sub-channels for a total of $S \cdot A$ angles each. For example, a $16 \times 16$ system with 320 MHz channels requires 256 complex elements for each of the 996 subcarriers. The 802.11 standard requires 8 bits for each real and imaginary component of the CBF, which results in 510 kB.  \vspace{-0.2cm}

\subsection{Challenges of 802.11 Beamforming Procedure}

The size of the beamforming feedback (BF) grows as $N_t \times \sum_{i=1}^{N_s} N_{ss, i} \times S$. This implies the following drawbacks:

$\bullet$ Feedback airtime increases with the number of \glspl{sta}, as each STA sends its \gls{bf} separately. Moreover, the number of decomposed angles and ultimately the size of the \gls{bf} depends on the number of antennas, and grows linearly with channel bandwidth, as discussed in Section \ref{sec:airtime_analysis}; 
    
$\bullet$ Computing and compressing the  through \gls{svd} and \gls{gr} operations imposes a significant computational load on beamformees, as discussed in detail in Section \ref{sec:comp_analysis}. This may impact resource-limited devices;

$\bullet$ \gls{gr} angle decomposition and \gls{bf} reconstruction introduce an additional error. This deteriorates the performance of the multi-user transmission, especially in scenarios with small inter-user separation where successful data recovery depends highly on accurate beamforming;

$\bullet$ The computational load at the STA and the feedback size cannot be modified according to application- and device-specific constraints. As next-generation Wi-Fi caters to heterogeneous devices and a wide range of performance requirements, it is critical to achieve this functionality.

\section{The \NW Framework}
In this section, we elaborate on the \NW framework. First, the system model and design challenges are outlined in Section \ref{sec:sys_model}. Next, the BOP is formulated and the heuristic solution is detailed in Sections  \ref{sec:optimization} and \ref{sec:BOP_solution}. Finally, the \NW model implementation and the customized training procedure are explained in Section \ref{sec:train}.

\subsection{The \NW DNN } \label{sec:sys_model}

\begin{figure}[!h]
    \centering
    \includegraphics[width=0.95\columnwidth]{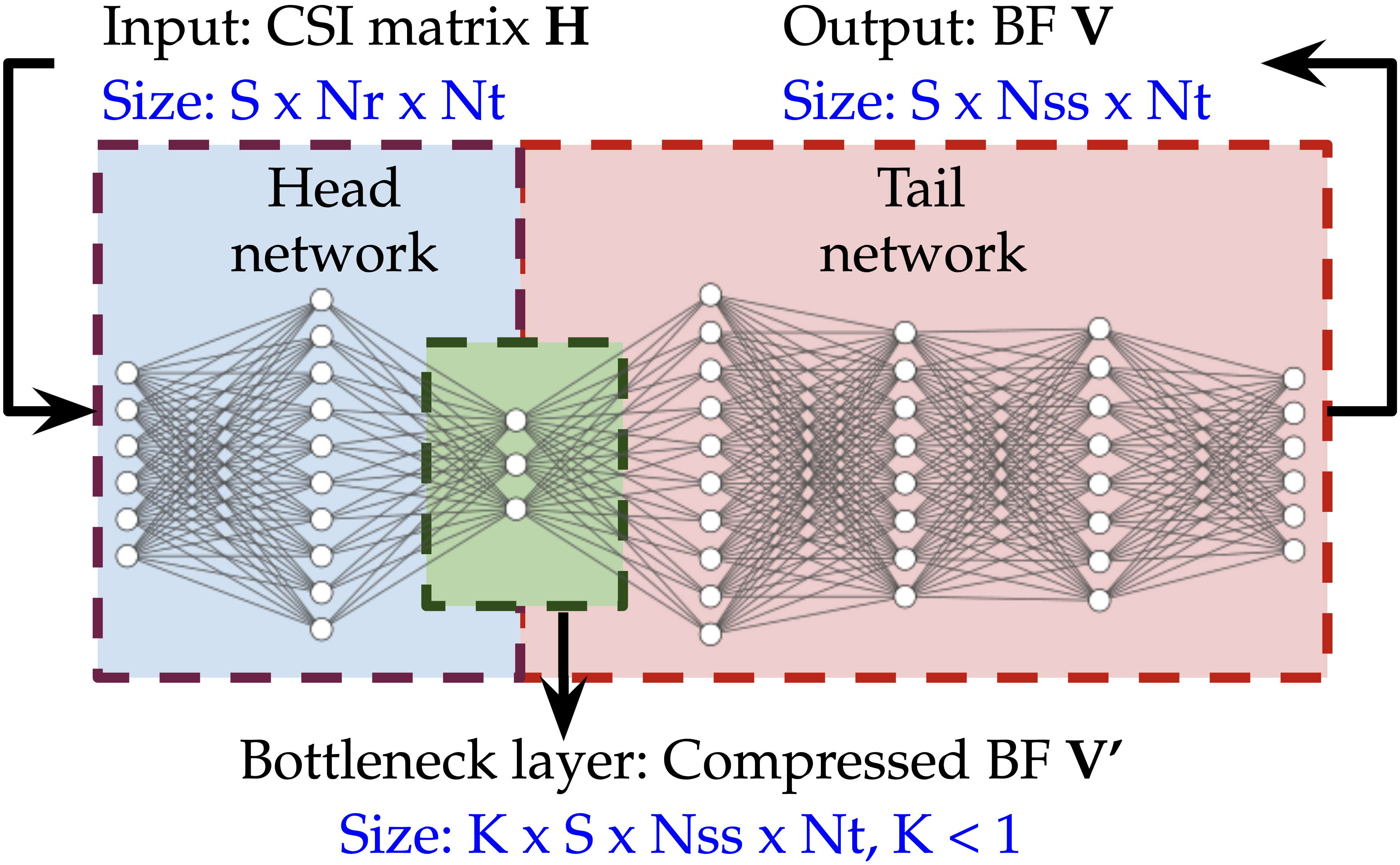}
    \caption{Head and tail networks in SplitBeam. Notice that we dropped the index $i$ in the mathematical notation for simplicity.\vspace{-0.2cm}}
    \label{fig:splitbeam_network}
\end{figure}

\NW trains a \gls{dnn} that maps the \gls{csi} matrix $\mathbf{H}_i$ to the \gls{bf} $\mathbf{V}_i$ in a supervised manner. To compress the \gls{bf} and transfer the \glspl{sta} computational load to the \gls{ap} (with higher computational capacity), 
we introduce  a ``bottleneck layer'' in the DNN as shown in Figure \ref{fig:splitbeam_network}. The bottleneck is an intermediate representation in the \gls{dnn} model which is ($K<1$ times) smaller than the model input $\mathbf{H}_{i}$. The bottleneck divides the \gls{dnn} into a \textit{head} and a \textit{tail} network, which are respectively executed on the STA and the \gls{ap}.

\begin{figure}[!ht]
    \centering
    \includegraphics[width=0.95\columnwidth]{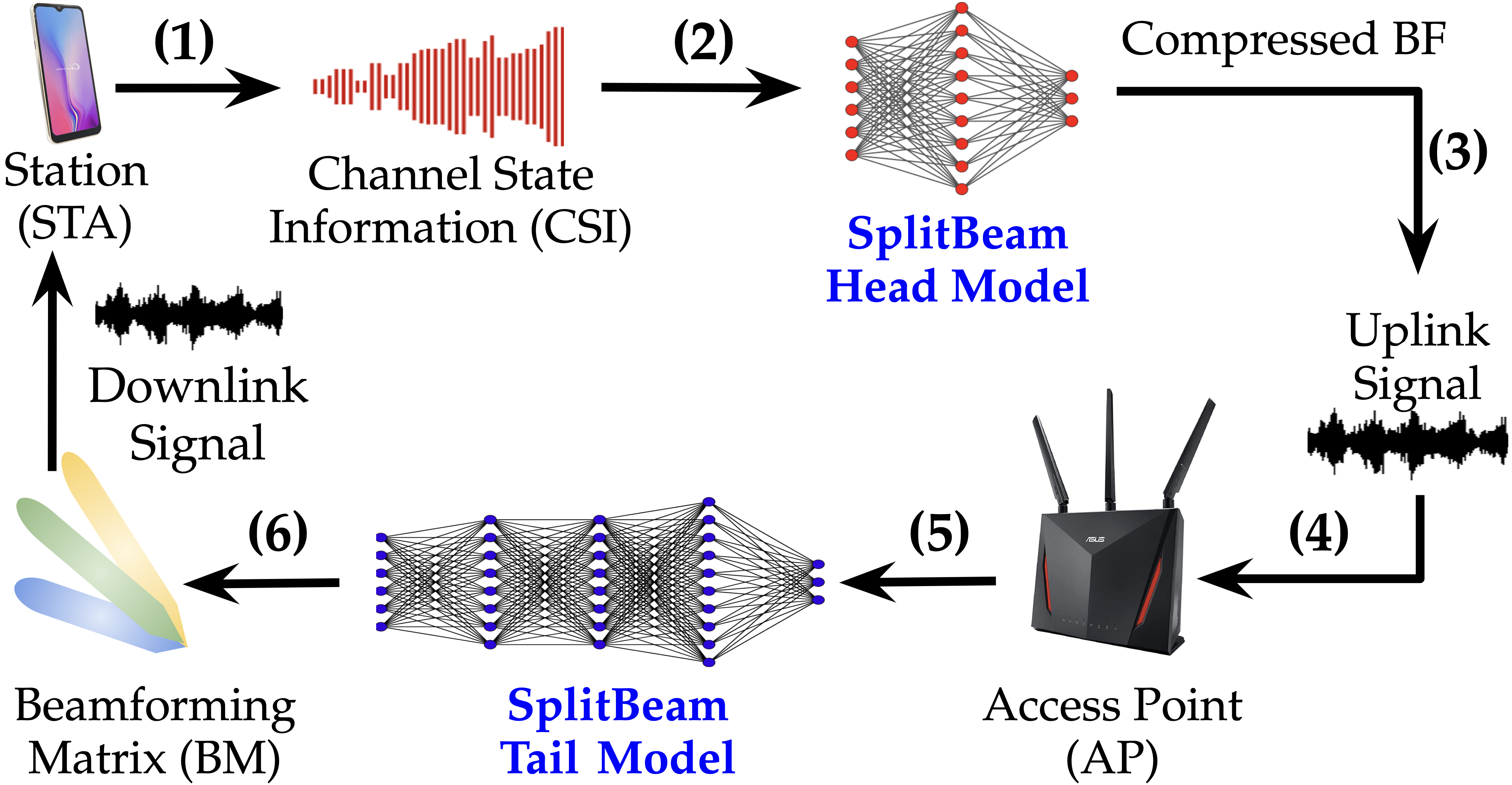}
    \caption{\NW beamforming feedback procedure.\vspace{-0.4cm}}
    \label{fig:system}
\end{figure}

An overview of \NW is shown in Figure \ref{fig:system}, where  \textbf{(1)} the estimated \gls{csi} matrices at STAs are fed to the head model \textbf{(2)} that is tasked to produce a compressed representation of the \gls{bf} denoted by $\mathbf{V}^\prime_i$ \textbf{(3)}.  The compressed \gls{bf} is sent to the \gls{ap} over the air \textbf{(4)}, where it is fed to the tail model \textbf{(5)} to reconstruct the \gls{bf} and generate the beamforming matrix\textbf{(6)}. 

\textbf{Remarks.}~The placement and size of the bottleneck ultimately determine the head network architecture, and thus (i) the \gls{sta} computational load, (ii) the \gls{bf} feedback size, and (iii) the beamforming accuracy. \textit{Indeed, there is a \textit{trade-off} between the complexity of the head model, the \gls{bf} compression rate, and the accuracy of inference}. While placing the bottleneck early on with a low number of nodes reduces the STA computation load and airtime overhead, it leads to a decrease in beamforming accuracy, which ultimately increases the \gls{ber}.
Therefore, the bottleneck placement and size must be adjusted according to the application-specific requirements. \vspace{-.1cm}

\subsection{Bottleneck Optimization Problem (BOP)} \label{sec:optimization} 
We model the original \gls{dnn} as a function $\mathcal{M}$ that maps the channel matrix $\mathbf{H}_i \in \mathbb{C}^{N_r \times N_t \times S}$ to the \gls{bf} $\mathbf{V}_i \in \mathbb{C}^{N_r \times N_t \times S}$ as $\mathcal{M}(\mathbf{H}; \boldsymbol \theta) : \mathbb{C}^{\mathbf{H}} \rightarrow \mathbb{C}^{\mathbf{V}}$, thorough $L$-layer transformations:
\begin{equation}
    \mathbf{r}_{j} = F_j(\mathbf{r}_{j-1}, \boldsymbol \theta_j) \hspace{0.5cm} 0 \le j \le L,
    \nonumber
\end{equation}
where $F_j(\mathbf{r}_{j-1}, \boldsymbol \theta_j)$ is the mapping carried out by the $j$-th layer and $j=0$ denotes the input layer. The vector $\boldsymbol{\theta} = \{\boldsymbol \theta_1, \ldots, \boldsymbol \theta_L\}$ defines the set of parameters of the \gls{dnn}. To devise the \textit{bottleneck}, we use an encoder-decoder like structure where the first $e$ layers of the \gls{dnn} is the encoder and the rest of the layers are the decoder. The encoder, called the head model $\mathcal{H}$, is placed from the input layer to the bottleneck $\mathcal{B}$. Next, the tail model $\mathcal{T}$ decompresses the encoded \gls{bf} to construct the \gls{bf} $\mathbf{V}_i$. The modified model can be written as
\begin{equation}\label{eq:dnn}
\mathcal{M}(\mathbf{H}; \boldsymbol \theta)=\left\lbrace \hspace{-0.5mm}
\begin{array}{ll}
\hspace{-1mm} \mathcal{H} = F_j(\mathbf{r}_{j-1}, \boldsymbol \theta_j) \text{,} \hspace{-1mm}&\hspace{-1mm} 0 \le j < e,\\
\hspace{-1mm} \mathcal{B} = F_e(\mathbf{r}_{e-1}, \boldsymbol \theta_e) \text{,} \hspace{-1mm}&\hspace{-1mm} j = e,\\
\hspace{-1mm}\mathcal{T}= F_j(\mathbf{r}_{j-1}, \boldsymbol \theta_{j}),&\hspace{-1mm} e+1 \le j \leq L.
\end{array}	\right.
\end{equation}

Let $L_i^{\mathcal{H}}(e,N)$ be the \gls{sta} $i$ overhead consists of three components: (i) the computational cost (i.e., the power consumption and memory required for executing the model), denoted by $L_i^c(e,N)$; (ii) the execution time for \gls{bf} compression through the head model, denoted by $T_i^{\mathcal{H}}(e,N)$; and (iii) the power consumption of transmitting the compressed \gls{bf} to the \gls{ap}, denoted by $L_i^{tx}(e,N)$. Also, $T_i^A(e,N)$ represents the compressed \gls{bf} feedback airtime. Finally, $T^{\mathcal{T}}(e,N)$ denotes the time required for reconstructing the \gls{bf} at the \gls{ap}. Notice that compression, decompression and airtime overhead depend on the placement $e$ and size of the bottleneck $N$. \smallskip

We define the BOP such that it minimizes the STA computation overhead and feedback airtime as
\begin{subequations}
\begin{align}
\underset{e, N}{\min} & \hspace{0.2cm} \sum_{i \in \mathcal{I}} \left(\mu_i^{\mathcal{H}} \hspace{0.1cm} L_i^{\mathcal{H}}(e,N) + (1 - \mu_i^{\mathcal{H}}) \cdot  \hspace{0.1cm} T_i^A(e,N)\right)  \label{eq:opt_obj} \\
    \textrm{s. t.} & \hspace{0.2cm} 0 <  \mu_i^{\mathcal{H}} < 1 , \hspace{0.2cm} i \in \mathcal{I} \\
    & \hspace{0.2cm}
    \mathrm{BER}_i \leq \gamma, \hspace{0.2cm} i \in \mathcal{I}\label{eq:opt_c1}  \\
    & \hspace{0.2cm} \max_{i \in \mathcal{I}} (T_i^{\mathcal{H}}(e,N) + T^{A}_i(e,N)) + T^{\mathcal{T}}(e,N) < \tau,  \label{eq:opt_c2}
\end{align}
\end{subequations}
where $\mu_i^{\mathcal{H}}$ parameterizes the importance of reducing the \glspl{sta} overhead versus the feedback airtime. In applications where STAs are resource-constrained, it is crucial to reduce the STAs load, i.e., $\mu_i^{\mathcal{H}}>\mu_i^A$. On the other hand, in dynamic propagation environments like crowded rooms, where the channel coherence time is short, high feedback airtime cannot be tolerated. Thus, reducing the feedback airtime must be prioritized, i.e., $\mu_i^{\mathcal{H}}<\mu_i^A$.
$\mathrm{BER}_i$ represents the bit error rate (BER) of client $i$. 
In this work, we measure the accuracy of the generated \gls{bf} at the \gls{ap} in terms of achievable \gls{ber} by the STAs. BER is the number of erroneous bits divided by the total number of transferred bits.  
Condition \eqref{eq:opt_c1} guarantees that the \gls{ber} experienced by each client does not exceed the maximum \gls{ber} threshold $\gamma$. Condition \eqref{eq:opt_c2} indicates that the maximum end-to-end delay of \gls{bf} cannot exceed the maximum tolerable delay denoted by $\tau$. In practice, these two conditions ensure that the bottleneck placement does not significantly impact the inference accuracy and latency. The maximum tolerable \gls{ber} and delay can be specified according to the requirements.\vspace{-0.2cm}

\subsection{Heuristic Procedure for Solving the BOP} \label{sec:BOP_solution}

The BOP is a particular instance of the extremely complex neural architecture search (NAS) problem \cite{xiong2019resource,NAS}.  Thus, we devise a heuristic algorithm to search for proper \NW hyperparameters that is specific to our context.
Specifically, to limit the search space, we take the following procedure:

\begin{enumerate}
    \item With the primary goal of minimizing the clients' computational load $L_i^{\mathcal{H}}$, we place the bottleneck layer immediately after the input layer (i.e., $e = 1$);
    \item To reduce the inference time at the \gls{ap}, $T^{\mathcal{H}}$, we consider only one layer for the tail network (i.e., $L=2$). Thus, the resulting \gls{dnn} is a 3-layer network comprising input, bottleneck and output layers;
    \item We adjust the size of the bottleneck layer according to the QoS requirements. Specifically, we consider a limited number of compression levels $K = \sfrac{\mathbf{V}^\prime_i}{\mathbf{H}_i}$, and consider the BER as our QoS metric. We start from the highest level of compression (lowest number of bottleneck nodes), and train the 3-layer \gls{dnn} with the CSI and corresponding $\mathbf{V}$ matrices dataset according to the customized procedure in Section \ref{sec:train}. Once trained, the generated BM by the DNN is used to estimate the BER at the STA by comparing the recovered and transmitted data bits, as explained in Section 5.2.1.
    \item If the desired BER cannot be achieved, the compression level is decreased. The new model is trained according to step (3) until the model is capable of meeting the BER constraint. If the compression level is the minimum, another layer is inserted after the bottleneck ($L = L + 1$), and the algorithm goes back to step 3.
\end{enumerate}
\textbf{Section \ref{sec:results} shows that the heuristic algorithm simplifies the search while maintaining acceptable performance.} \vspace{-0.2cm}

\subsection{\NW Model Training} \label{sec:train}

Since $\mathbf{H}$ and $\mathbf{V}$ are  complex matrices,  we decouple real and complex components in the matrices and treat them as double-sized real matrices. For each of our datasets, we split a dataset into training, validation, and test splits with 8:1:1 ratio. \NW is trained offline for various network configurations and does not require retraining. The STAs select the proper trained DNN according to the network configuration information acquired from the \gls{ndp} preamble. 

\subsubsection{Loss Function}
Our goal is to deploy exactly the same model for each STA without fine-tuning its parameters to its environment. Notice that the training process is done offline (i.e., on a single computer). Given a channel matrix $\mathbf{H}_{i}$, our \gls{dnn} model $\mathcal{M}$ estimates the corresponding \gls{bf} $\mathbf{V}_{i}$, i.e., $\mathbf{V}_{i} = \mathcal{M}(\mathbf{H}_{i}, \boldsymbol \theta)$. We formulate the loss function $\mathcal{L}$ as follows
\begin{equation}
    \mathcal{L} = \frac{1}{b} \sum_{j=1}^{b} \sum_{i=1}^{N_{s}} \frac{\left( \mathcal{M}(\mathbf{H}_{i}^{j}, \boldsymbol \theta) - \mathbf{V}_{i}^{j} \right)^2}{\left\| \mathbf{V}_{i}^{j} \right\|_{1}},
    \label{eq:loss}
\end{equation}
\noindent where $b$ indicates training batch size and $\|\cdot\|_{1}$ represents L1-norm. 
$\mathbf{H}_{i}^{j}$ and $\mathbf{V}_{i}^{j}$ indicate the $j$-th channel matrix and \gls{bf} for STA $i$, respectively.
By minimizing the loss in \eqref{eq:loss}, we optimize the parameters $\boldsymbol \theta$ of our \gls{dnn} model $\mathcal{M}$. We use stochastic gradient descent (SGD) and Adam~\cite{kingma2015adam} to train the synthetic and experimental datasets, respectively. Unless specified, we train models for 40 epochs, using the training split in the dataset with batch size of 16 and the initial learning rate of $10^{-3}$. The learning rate is decreased by a factor of 10 after the end of 20th and 30th epochs. Using the validation split in the dataset, we assess the model in terms of achieved BER at the end of every epoch and save the best parameters $\boldsymbol \theta^{*}$ such that achieve the lowest BER for the validation split.
The trained model is assessed with the best parameters for the held-out test split in the dataset and report the test BER. 

\subsubsection{Difference with Autoencoders}Although an \gls{ae} is similar in terms of model architecture, its training objective is different. \glspl{ae} are trained to reconstruct its input in an unsupervised manner (e.g., to estimate $\hat{\mathbf{V}}_{i}$ given $\mathbf{V}_{i}$) as done in \cite{sangdeh2020lb}. Conversely, we train a task-specific model in \emph{a supervised fashion} to estimate \gls{bf} $\mathbf{V}_{i}$ given a channel matrix $\mathbf{H}_{i}$. \vspace{-0.1cm}

\subsection{Complexity Analysis and Compression Rate} \label{sec:complexity}
\subsubsection{Computational Overhead}\label{sec:comp_analysis}
The complexity of the \gls{svd} operation for decomposing the \gls{bf} $\mathbf{V}$ in 802.11 is $\mathcal{O}( (4N_t N_r^2 + 22  N_t^3)S )$, according to \cite{golub1996matrix}. The \gls{bf} is further transformed into a set of angles using the Givens rotation (GR) matrix multiplication which has a complexity of $\mathcal{O}( N_t^3 N_r^3 S)$ \cite{perahia2013next}. Conversely, the complexity of \NW is $\mathcal{O}( K N_t^2 N_r^2 S^2)$, where $K < 1$ denotes the head model's compression level.


\begin{figure}[!h]
    \centering
    \vspace{-0.2cm}
    \includegraphics[width=0.7\columnwidth,angle=-90]{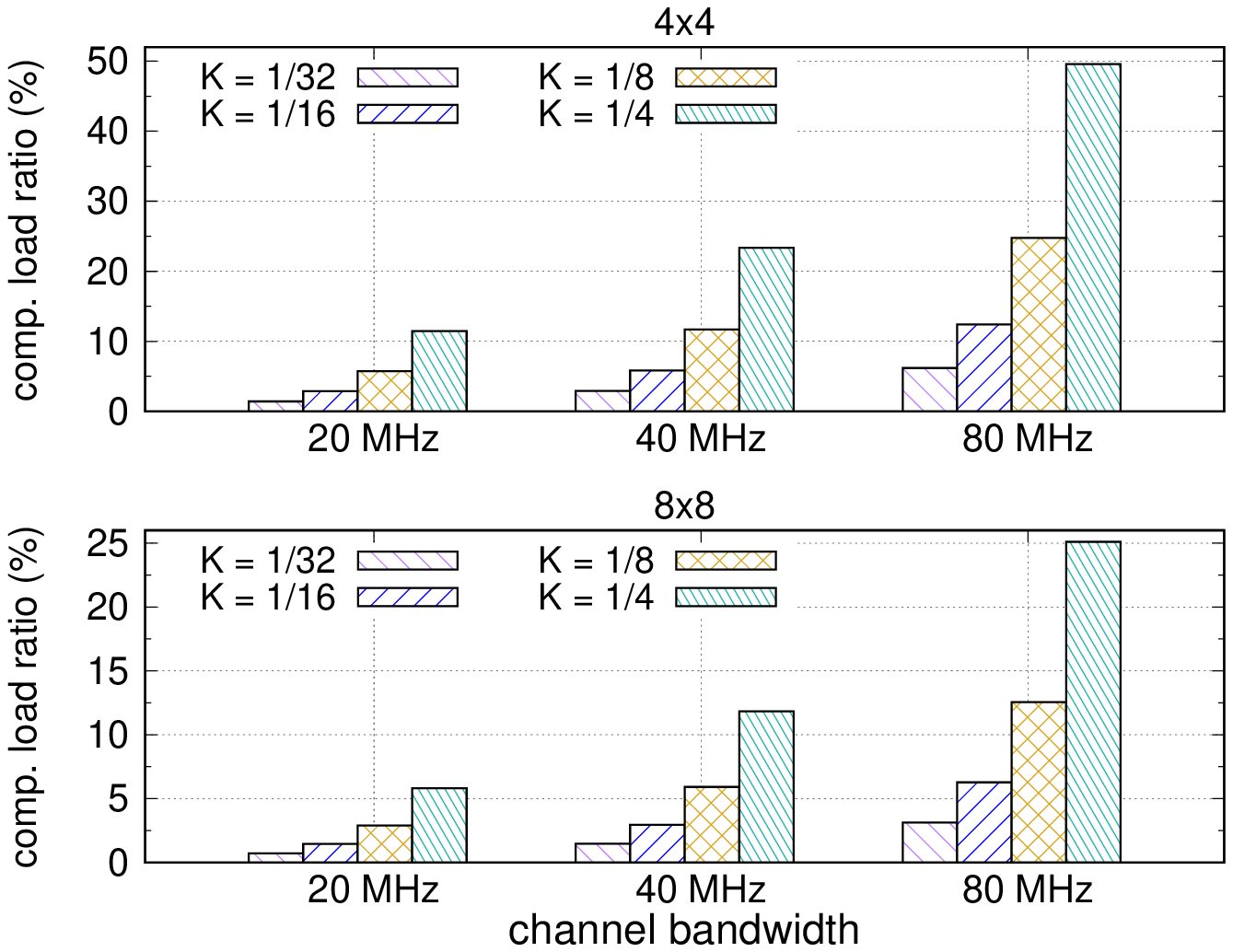}
    \caption{The ratio of the computational load using \NW to 802.11 with $N_{ss,i} = 1$ and various compression levels $K$ and channel widths.\vspace{-0.2cm}}
    \label{fig:COMP_COMPLEXITY}
\end{figure}

Figure \ref{fig:COMP_COMPLEXITY} shows the ratio of the number of floating-point operations (FLOP) required for compressing the \gls{bf} using \NW to 802.11 compression technique. The ratio is calculated by $\sfrac{X}{Y}\times 100$ where $X$ and $Y$ denote the number of floating points operation in \NW and legacy Wi-Fi protocol, respectively. 
The comparison is performed for different MU-MIMO orders and a various number of subcarriers, as computed through a MATLAB program. We can see that \NW noticeably reduces the computational load of \gls{sta}, especially as the number of antennas and/or \glspl{sta} increases. At 80 MHz, \NW with $K=\sfrac{1}{8}$ decreases 75\% and 87\% of the \gls{sta}'s load in $4 \times 4$ and $8 \times 8$ systems. \textbf{On average, \NW improves computation by \textbf{73\%}. We show in Section \ref{sec:results} that \NW with $K = \sfrac{1}{8}$ keeps the \gls{ber} within 87\% of 802.11}.

\subsubsection{Airtime Overhead}\label{sec:airtime_analysis}
In 802.11, the size of the compressed \gls{bf} report is $BMR = 8 \times N_t  + N_a \times S \times (b_{\phi}+ b_{\psi})/2$ where $N_a$ denotes the number of Givens angles \cite{perahia2013next}. Notice that $b_{\phi}$ and $ b_{\psi}$ are the number of bits required for the angle quantization \cite{gast2013802}. 
Therefore, the 802.11 compression ratio can be written as 
\begin{equation}
    CR = \frac{BMR}{S\times N_t \times N_r \times b} \vspace{3pt},\label{eq:com_ratio}
\end{equation}
where $b=16$ is the number of bits required for transmitting channel information over each subcarrier.  Conversely, the compression rate of \NW is  $K$. Notice that it is  constant and does not grow with the size of the channel matrix.

\begin{figure}[!h]
    \centering
    \vspace{-0.3cm}
    \includegraphics[width=0.7\columnwidth,angle=-90]{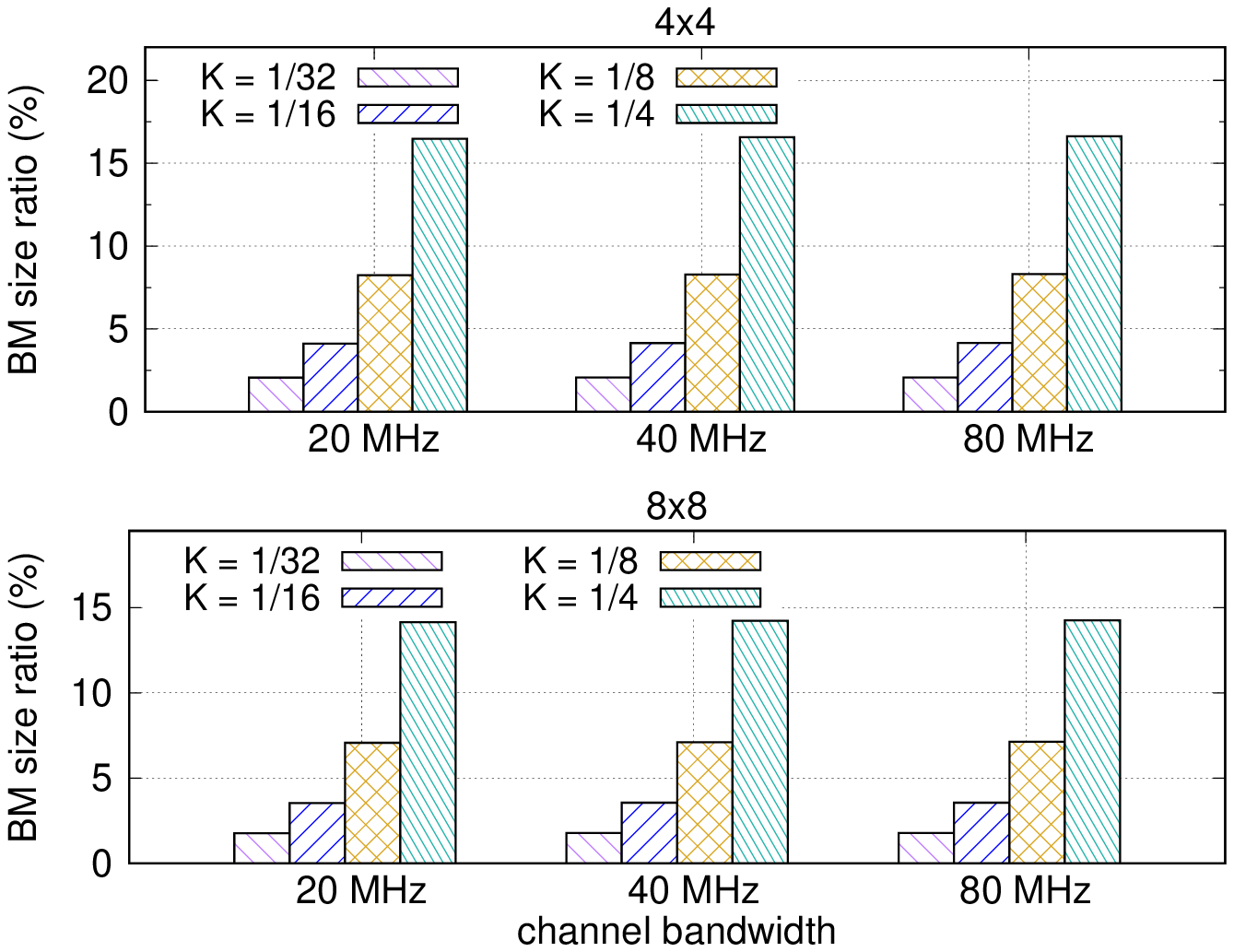}
    \caption{The ratio of the size of \gls{bf} using \NW to 802.11 with $N_{ss,i} = 1$ and various compression levels $K$ and channel widths.     \vspace{-0.4cm}}
    \label{fig:air_overhead}
\end{figure}

Figure \ref{fig:air_overhead} depicts the impact of \NW in reducing the airtime overhead. The bars show the ratio of the size of the compressed \gls{bf} of \NW to the angle decomposition technique in 802.11. \NW has a significant impact at higher-order \gls{mumimo} configurations. For example, \NW reduces the size of the feedback overhead by 91\% and 93\% in  $4 \times 4$ and $8 \times 8$ configurations with 80 MHz channel. \textbf{On average, \NW reduces the airtime overhead by \textbf{75\%} with respect to 802.11.} 

\section{Experimental Evaluation}

We first introduce the experimental \gls{mumimo} \gls{csi} extraction method along with the measurement campaigns for building the datasets in Section \ref{setup}. Next, we detail organization of the dataset, the training and testing procedure of the \NW using the collected datasets in Section \ref{training}.\vspace{-0.2cm}

\subsection{Experimental Setup}\label{setup}

We designed our testbed with commercially-available off-the-shelf Wi-Fi devices to collect real-world datasets. As explained in Section \ref{sec:train}, training the \NW requires downlink \gls{csi} that is measured at the \gls{sta} side, in addition to the corresponding \gls{bf}. The measurements are carried over different network configurations to verify the performance of \NW as the number of subcarriers and antennas and \glspl{sta} increases. We consider different propagation environments to test how \NW generalizes. \smallskip

\begin{figure}[!t]
    \centering
    \includegraphics[width=0.95\columnwidth, trim={0cm .3cm  0cm 0cm },clip]{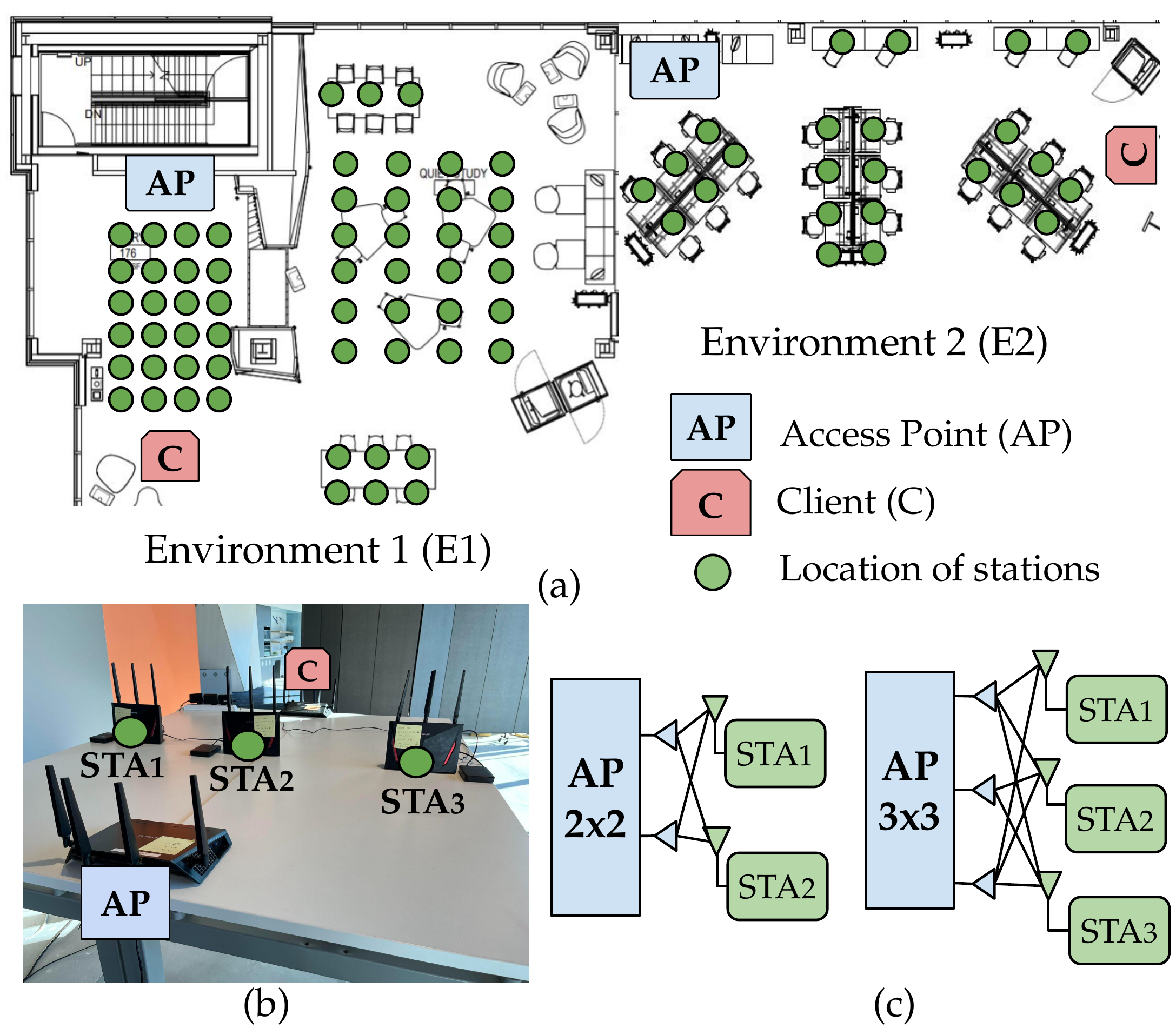}
    \caption{Experimental setup for MU-MIMO CSI data collection: (a) different environments; (b) network elements; (c) network configurations.\vspace{-0.4cm}}
    \label{fig:testbed}
\end{figure}

\noindent\textbf{5.1.1: CSI Extraction.}~In commercial Wi-Fi chipsets, \gls{csi} data is estimated through pilot symbols. Being computed at the \gls{phy}, \gls{csi} is not accessible by the end-user through normal \glspl{nic}. Thus, we have used Nexmon \gls{csi} \cite{nexmonCSI}, the state-of-the-art \gls{csi} extraction tool to collect \gls{csi} measurements using Asus RT-AC86U 802.11ac Wi-Fi routers as STAs. The extraction tool is compatible with the \gls{vht} mode, defined by IEEE 802.11ac, working with bandwidth up to 80 MHz. Each CSI sample results in complex-valued channel information per subcarrier for each transmit-receive antennas pair. A Netgear R7800 Wi-Fi router with a Qualcomm Atheros chipset is used as \gls{ap}.   An example of the experimental setup realization is shown in Figure \ref{fig:testbed}(b). In a real-world scenario, \NW relies on already existing channel estimations at the STAs. However, since the Nexmon tool is configured for reading CSI samples only on data packets, we established a Wi-Fi link between the \gls{ap} and another Netgear R7800 Wi-Fi router (as the client) to generate the data packets that provides the ASUS \glspl{sta} the opportunity to extract CSI.

\subsection{Data Collection and Model Training}\label{training}

Packets are transmitted with a rate of 1000 packets/second through $N_t$ antennas with a fixed modulation and coding scheme. Thus, a new \gls{csi} is generated every $T = 10^{-3}$ s.  To evaluate the capability of \NW in generalizing to different environments, we performed \gls{csi} measurements in two environments $E1$ and $E2$.  We carefully picked the environment with exclusive furniture arrangements, size, and construction material to ensure that the target environments are mutually exclusive from the source environment in terms of propagation characteristics. Specifically, $E1$ has fewer reflectors and human traffic, while $E2$ is furnished with more furniture (multipath) and is imposed to higher human traffic. Figure \ref{fig:testbed}(a) displays the positions of the \gls{ap} and \glspl{sta} in the different environments. 
To capture the impact of inter-user distance on CSI data, the \glspl{sta} are placed in different distances from each other (15-60 cm). Also, users are located at different distances from the \gls{ap} (0.5-6 m). The green dots in Figure \ref{fig:testbed}(a) depicts the location of points where \glspl{sta} are located for data collection.\smallskip

\noindent\textbf{5.2.1: Datasets.}~
We consider $2 \times 2$ and $3 \times 3$ scenarios, where the \gls{ap} with $N_{t} = 2,3$ antennas simultaneously serves two and three STAs. 
The network configurations are shown in Figure \ref{fig:testbed}(c), where each \gls{sta} device supports one spatial stream, i.e., $N_{ss,i}=1$ for $i \in \mathcal{I}$. Moreover, the Nexmon tool enabled us to collect 802.11ac channel measurements at 5 GHz with 20, 40 and 80 MHz bandwidth over $|\mathcal{S}| = 56$, $|\mathcal{S}|=114$ and $|\mathcal{S}| = 242$ subcarriers, to assess the performance of the \NW at higher channel bandwidth. The measurements are repeated several times with a time interval of at least 4 hours in between measurements. To capture the impact of human blockage and reflection, the datasets are collected during working and non-working days. We collected 10,000 \gls{csi} samples per configuration, channel width and environment. In total, 12 datasets with 120,000 data samples were extracted to train, evaluate and test the \NW. Tables \ref{tab:datasets} shows the list of collected experimental datasets. \textbf{We pledge to release, along with the code, the collected dataset for full reproducibility.}

\begin{table}[!h]
\footnotesize
\centering
\begin{tabular}{ccc|ccc|}
\cline{4-6}
\multicolumn{3}{c|}{}                                                                                                   & \multicolumn{3}{c|}{\textbf{Config.}}                                                                                                                   \\ \hline
\multicolumn{1}{|c|}{\textbf{Type}}                  & \multicolumn{1}{c|}{\textbf{BW (MHz)}}           & \textbf{Env.} & \multicolumn{1}{c|}{\textbf{$2 \times 2$}} & \multicolumn{1}{c|}{\textbf{$3 \times 3$}} & \textbf{$4 \times 4$} \\ \hline
\multicolumn{1}{|c|}{}                               & \multicolumn{1}{c|}{}                            & $E1$          & \multicolumn{1}{c|}{$\mathcal{D}_1$}                               & \multicolumn{1}{c|}{$\mathcal{D}_2$}                               & -                                             \\ \cline{3-6} 
\multicolumn{1}{|c|}{}                               & \multicolumn{1}{c|}{\multirow{-2}{*}{20}}        & $E2$          & \multicolumn{1}{c|}{$\mathcal{D}_3$}                               & \multicolumn{1}{c|}{$\mathcal{D}_4$}                               & -                                             \\ \cline{2-6} 
\multicolumn{1}{|c|}{}                               & \multicolumn{1}{c|}{}                            & $E1$          & \multicolumn{1}{c|}{$\mathcal{D}_5$}                               & \multicolumn{1}{c|}{$\mathcal{D}_6$}                               & -                                             \\ \cline{3-6} 
\multicolumn{1}{|c|}{}                               & \multicolumn{1}{c|}{\multirow{-2}{*}{40}}        & $E2$          & \multicolumn{1}{c|}{$\mathcal{D}_7$}                               & \multicolumn{1}{c|}{$\mathcal{D}_8$}                               & -                                             \\ \cline{2-6} 
\multicolumn{1}{|c|}{}                               & \multicolumn{1}{c|}{}                            & $E1$          & \multicolumn{1}{c|}{$\mathcal{D}_9$}                               & \multicolumn{1}{c|}{$\mathcal{D}_{10}$}                            & -                                             \\ \cline{3-6} 
\multicolumn{1}{|c|}{\multirow{-6}{*}{Real}}         & \multicolumn{1}{c|}{\multirow{-2}{*}{80}}        & $E2$          & \multicolumn{1}{c|}{$\mathcal{D}_{11}$}                            & \multicolumn{1}{c|}{$\mathcal{D}_{12}$}                            & -                                             \\ \hline
\rowcolor[HTML]{EFEFEF} 
\multicolumn{1}{|c|}{\cellcolor[HTML]{EFEFEF}Synth.} & \multicolumn{1}{c|}{\cellcolor[HTML]{EFEFEF}160} & MATLAB        & \multicolumn{1}{c|}{\cellcolor[HTML]{EFEFEF}$\mathcal{D}_{13}$}    & \multicolumn{1}{c|}{\cellcolor[HTML]{EFEFEF}$\mathcal{D}_{14}$}    & $\mathcal{D}_{15}$                            \\ \hline
\end{tabular}
\caption{Datasets collected during our data collection campaign.\vspace{-.2cm}}
\label{tab:datasets}
\end{table}

We have used the MATLAB WLAN toolbox to generate a dataset at 160 MHz. This is because our experimental setup did not allow us to collect CSI at 160 MHz and with $4 \times 4$ MIMO. We have used \textit{wlanTGacChannel} function that filters an input signal through an 802.11ac multipath fading channel. The multi-user channel consists of independent single-user MIMO channels between the AP and spatially separated stations.  Each user estimates its own channel using the received NDP signal and computes the CSI. The delay profile ``Model-B'' has been considered which respectively consists of 9 channel taps and 2 channel clusters. Dataset $\mathcal{D}_{13}$-$\mathcal{D}_{15}$ each contain 10,000 data points. To remove noise and unwanted amplification, the \gls{csi} elements are normalized by the mean amplitude over all subcarriers. In addition, to remove the noise a $n$-point moving median window with $n=10$ is used to smooth out the noisy data. In addition, we noticed that in some instances, \gls{csi} packets are dropped by some \glspl{sta}. Therefore, using the packets sequence number, the data collected from different devices are aligned to ensure that each \gls{csi} element collected over different \glspl{sta} represents the same time and frequency domain channel measurements for seamless beamforming. \smallskip

\noindent\textbf{5.2.2: BER Computation.}~A key issue is that BER extremely depends on noise and fading levels, which makes it challenging to isolate the BER caused by the DNN compression. For this reason, and the sake of repeatability, we have used a MATLAB-based program to compute the BER corresponding to a given DNN compression. We have set the total number of transmit antennas to the sum of all the used spatial streams, so that no space-time block coding (STBC) or spatial expansion is needed at the \gls{ap}. Moreover, no channel coding  is considered, unless otherwise specified. The BER measurement procedure for each collected CSI data point $j$ is as follows: (1) we randomly generate bits modulated with 16-QAM that are used as payload to generate $N_s$ standard-compliant 802.11 frames $\mathbf{F}^{j} = [\mathbf{F}^{j}_{1} \cdots \mathbf{F}^{j}_{N_s}$] for $N_s$ receiving STA each served with one spatial stream. Denoting the $j$-th CSI value collected from the $i$-th STA as $\mathbf{H}_{i}^{j}$, (2) we run the \NW trained head and tail models $\mathcal{M}_i$ for each user to compute the $\mathbf{V}_{i}^{j}$ values corresponding to the $\mathbf{H}_{i}^{j}$ inputs; (3) we compute $\mathbf{H}_\text{EQ}^{j} = [\mathbf{V}_{1}^{j}, \cdots, \mathbf{V}_{N_s}^{j}]$ as described in Section \ref{sec:prelim_beamforming};  (4) we use zero-forcing (ZF) beamforming to calculate $\mathbf{W}^{j}$ as
\begin{equation}
    \mathbf{W}^{j} = \mathbf{H}_{EQ}^{j} \cdot (\mathbf{H}^{j\dagger}_{EQ} \cdot\mathbf{H}_{EQ}^{j})^{-1}.\nonumber
\end{equation}
(5) The received packets are generated as $\mathbf{Y}^{j} = \mathbf{H}_i^{j} \hspace{3pt} \mathbf{W}_i^{j}\hspace{3pt} \mathbf{F}_i^{j} + \mathbf{N}_i$, where $\mathbf{N}_i$ is Gaussian white noise. Finally, (6) the packets are demodulated, and the recovered bits are compared with the transmitted bits to calculate the BER.\smallskip

\noindent\textbf{5.2.3: Model Training and Testing.}~For each CSI measurement dataset in Table \ref{tab:datasets} the corresponding \gls{bf} dataset is generated using \gls{svd}. Next, \NW is trained to map the CSI measurement to \gls{bf} in a supervised manner, as detailed in Section \ref{sec:train}. We used  $K = \sfrac{1}{32}$, $\sfrac{1}{16}$, $\sfrac{1}{8}$, and $\sfrac{1}{4}$ as compression levels. A model is trained for each scenario, by using 80\% and 10\% of each dataset for training and validation. We designed two testing procedures: (i) \textit{single-environment} test, where the trained model is tested on the remaining 10\% of its dataset; (ii) \textit{cross-environment} test, where the model is tested on its counterpart dataset from the other environment. For example, let $\mathcal{M}_1$ be the model that is trained with $\mathcal{D}_1$ which is a $2 \times 2$ dataset collected in $E1$ at 20 MHz channel width. The model $\mathcal{M}_1$ is tested on: (i) $\mathcal{D}_1$ test-split that was held out during the training; (ii) $\mathcal{D}_3$ which is the dataset with the same configuration and channel width  in $E2$.

\section{Experimental Results}\label{sec:results}

We first compare the compression rate of \NW with respect to 802.11 and state-of-the-art LB-SciFi \cite{sangdeh2020lb} in Section \ref{sec:80211}. LB-SciFi uses an autoencoder (AE) to compress the angles generated by the 802.11 \gls{bf} compression algorithm. Finally, we evaluate the \NW generalization and efficiency results in Section \ref{sec:fpga}.\vspace{-0.1cm}

\subsection{Comparison with 802.11 and LB-SciFi}\label{sec:80211}

\begin{figure}[!ht]
    \centering
    \includegraphics[width=.7\columnwidth,angle = -90]{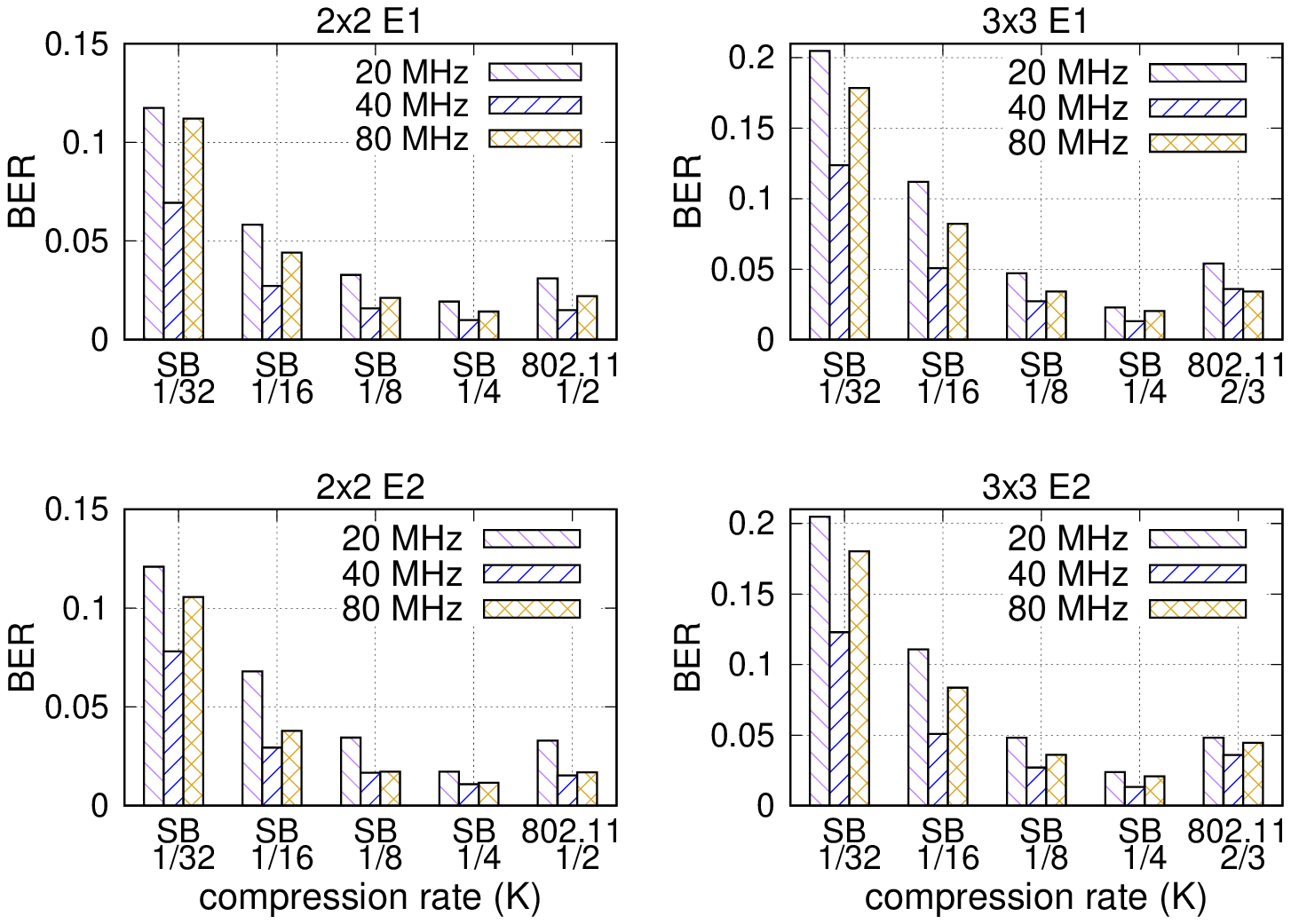}
    \caption{BER as a function of compression rate ($K = V'/H$): \NW(SB) vs 802.11; $2 \times 2$ and $3 \times 3$ MU-MIMO configurations in $E1$ and $E2$. Notice that in IEEE 802.11 $K\simeq 1/2$ and $2/3$ in  $2 \times 2$ and $3 \times 3$ configurations as the compression rate depends on network parameters as Equation \eqref{eq:com_ratio}. \vspace{-0.3cm}}
    \label{fig:BER_K_all}
\end{figure}

Figure \ref{fig:BER_K_all} depicts the trade-off between \gls{bf} compression rate ($K=\mathbf{V}'/\mathbf{H}$) and the incurred \gls{ber} with respect to 802.11. It can be seen that as the compression rate decreases and the \gls{bf} gets more compressed, the \gls{ber} increases.  However, we observe that the size of the feedback is much higher in 802.11. It can be seen that the \NW with a compression rate of $K=\frac{1}{8}$ achieves a \gls{ber} close to -- in some instances lower than -- the legacy Wi-Fi protocol while its size of \gls{bf} is respectively 4 and 5 times smaller in $2 \times 2$ and $3 \times 3$ configurations.

\begin{figure}[!ht]
    \centering
    \includegraphics[width=.65\columnwidth, angle=-90]{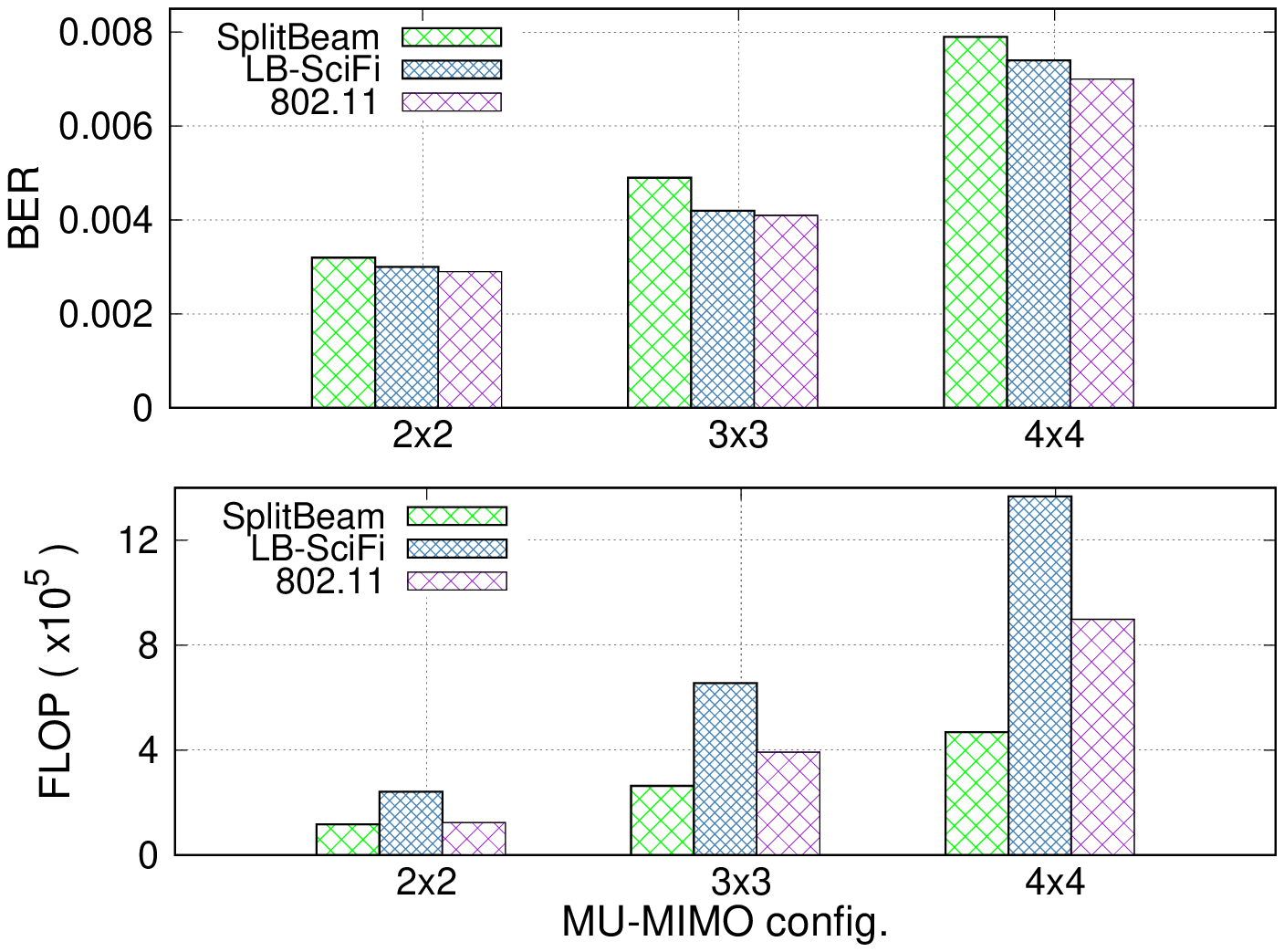}
    \caption{BER and STAs computational load for synthetic datasets with 160 MHz channel and compression rate of $K=\frac{1}{8}$.\vspace{-0.4cm}}
    \label{fig:simulation}
\end{figure}

Figure \ref{fig:simulation} shows achievable \gls{ber} and computational load, in terms of number of floating point operations (FLOP), for 160 MHz Wi-Fi transmissions (datasets $\mathcal{D}_{13}$ -- $\mathcal{D}_{15}$).  For these results, we used binary convolutional coding (BCC) with a code rate of 1/2. \NW achieves BER close to legacy 802.11 standard and LB-SciFi, which is the desired level. However, both 802.11 and LB-SciFi require \gls{svd} and \gls{gr} operations that impose high computational load on clients to achieve this performance. Figure \ref{fig:simulation} shows that \textbf{\NW reduces the computational load by 65\% and 45\% with respect to 802.11 and LB-SciFi} as it directly compresses the CSI matrix. When \NW is combined with channel coding, the BER is reduced significantly. Moreover, higher MU-MIMO orders are more sensitive to \gls{bm} estimation error.

\begin{figure}[!ht]
    \centering
    \includegraphics[width=1\columnwidth, trim={3.3cm 8.9cm  3.3cm 9cm },clip]{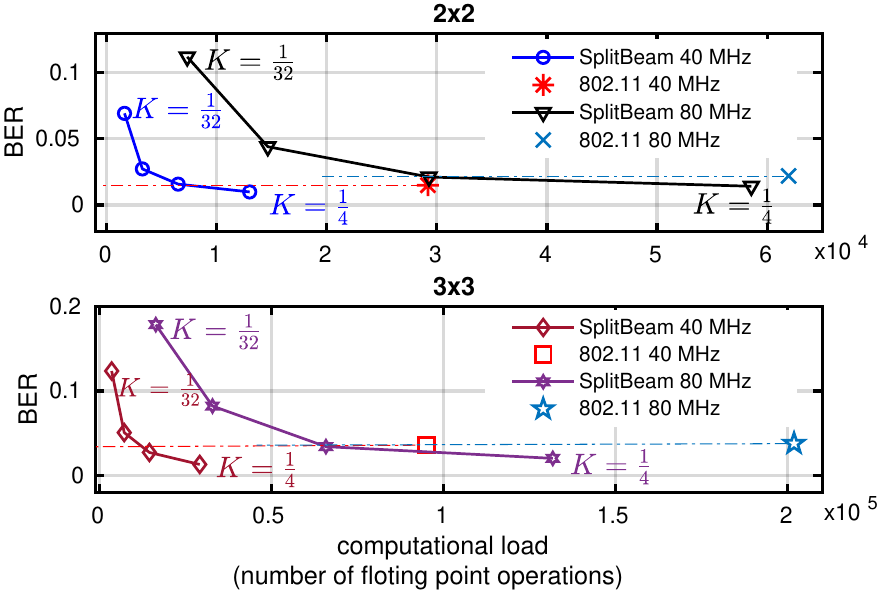}
    \caption{\gls{ber} as a function of computational STAs computational load: \NW vs 802.11; for $2 \times 2$ and $3 \times 3$ MU-MIMO configurations.\vspace{-0.3cm}}
    \label{fig:legacy_vs_SplitBeam_cross}
\end{figure}

Figure \ref{fig:legacy_vs_SplitBeam_cross} compares the achievable \gls{ber} of the \NW with 802.11 protocol as a function of computational load. We observe that \NW  maintains the \gls{ber} of the \gls{sta} close to the 802.11 protocol while imposing a considerably lower computational load on the users. Specifically, \textbf{\NW decreases the computation load by 70\% with respect to 802.11 while maintaining the same BER value of 0.02}.  We notice that the improvement given by \NW is more prominent when the number of antennas increases. For instance, \NW with $K = \frac{1}{8}$ decreases STAs' computational load respectively by 52\% and 68\% for $2 \times 2$ and $3 \times 3$ MU-MIMO.

\subsection{Model Generalization and Efficiency}\label{sec:fpga}

\begin{figure}[!ht]
    \centering
    \includegraphics[width=.65\columnwidth, angle = -90]{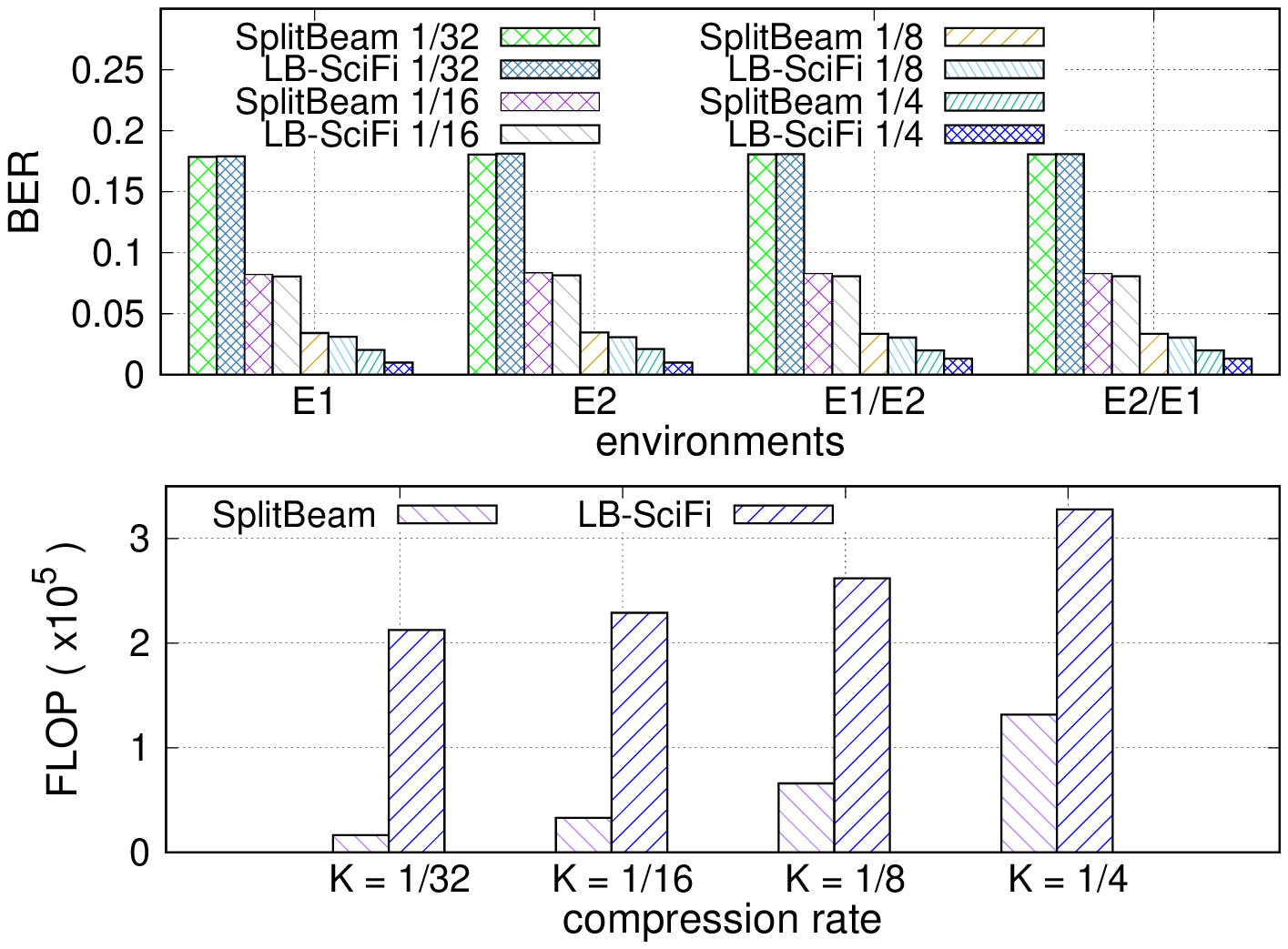}
    \caption{\gls{ber} and computational load for \NW vs LB-SciFi;
    single- and cross-environment test for $3 \times 3$ MU-MIMO with 80 MHz channel widths. Notice that BER is shown for $K=\frac{1}{8}.$\vspace{-0.2cm}}
    \label{fig:AE_vs_SplitBeam_cross}
\end{figure}

\begin{figure}[!h]
    \centering
    \includegraphics[width=.7\columnwidth, angle=-90]{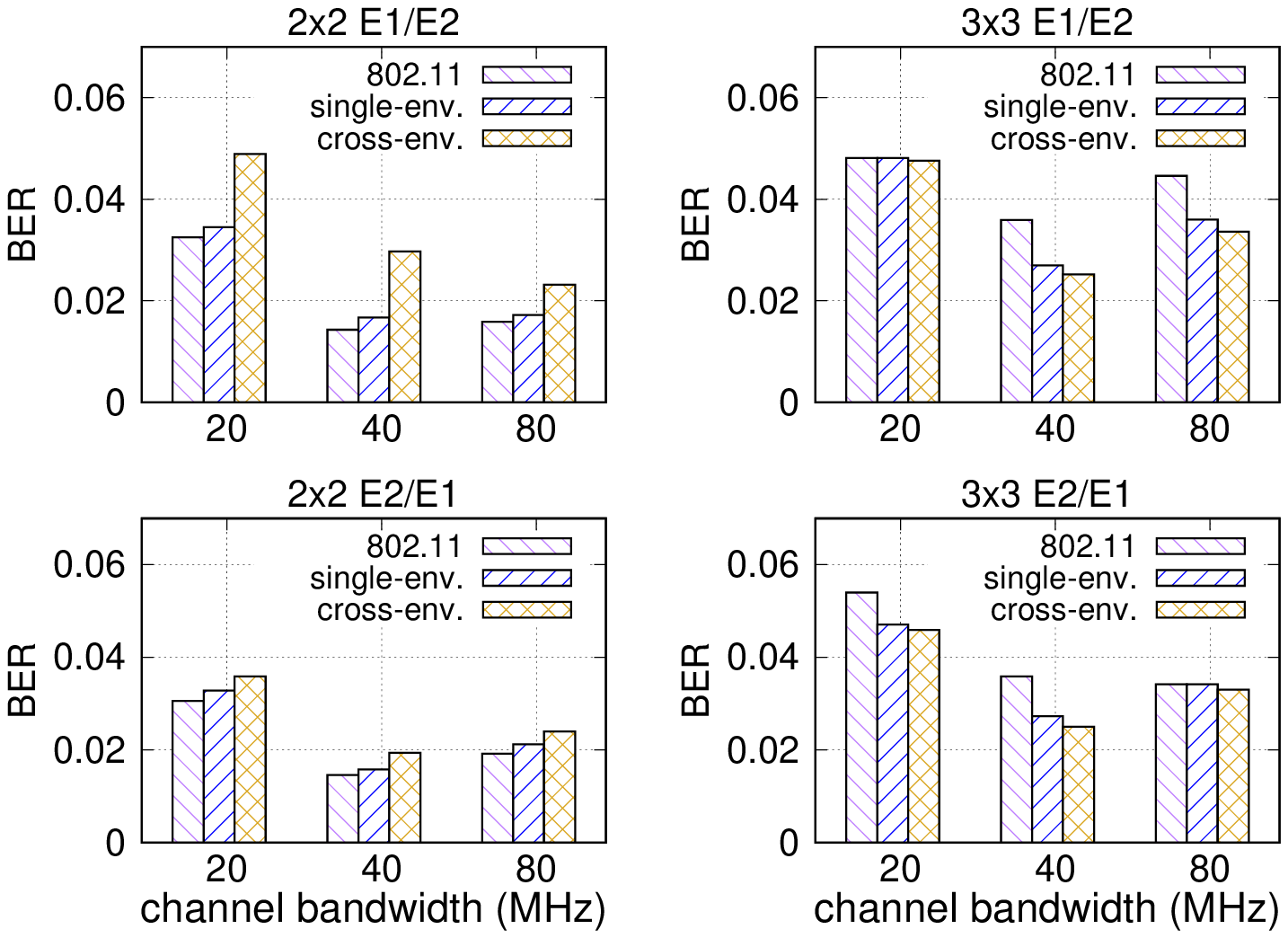}
    \caption{Cross-environment test: BER for $3 \times 3$ and $2 \times 2$ MU-MIMO configurations with compression rate of $K=\frac{1}{8}$. Note that $X/Y$ indicates that the model is trained in $X$ and tested in $Y$.\vspace{-0.3cm}}
    \label{fig:BER_K_cross}
\end{figure}

 Figure \ref{fig:AE_vs_SplitBeam_cross} shows the \gls{ber} and computational complexity for $3\times3$ \gls{mumimo} configuration with 80 MHz channel bandwidth in different environments. In the cross-environment setting $E1/E2$ ($E2/E1$), the models are trained and validated with the data collected from $E1$ ($E2$) and tested with the data from $E2$ ($E1$).  Although LB-SciFi achieves the same level of compression and BER as \NW (slightly lower in some cases), its computational load on \glspl{sta} is much higher than \NW. Specifically, on average, \textbf{\NW improves the computational load by  78\% with respect to LB-SciFi, while maintaining similar BER}. This is since 
 our approach compresses the estimated channel directly thus offloading devices' overhead significantly. Figure \ref{fig:BER_K_cross} further depicts the capability of \NW to generalize to untrained environments. We observe that in most cases, the cross-environment test has a performance close to the single-environment test.  Interestingly, we observe that the BER is usually lower when models are trained in $E2$ and tested in $E1$. This is because $E2$ has a more complex propagation profile than $E1$, i.e., more reflectors and presence of people. Thus, $E2$-trained models are more comprehensive and thus generalizing better. 

Table \ref{tab:bottleneck} investigates the trade-off among the head model complexity, the size of the \gls{bf}, and BER, as a function of the bottleneck structure in a $2 \times 2$ MIMO network. We compare the performance of the 3-layer \gls{dnn} -- designed and trained using the procedure explained in \ref{sec:BOP_solution} -- with more complex \glspl{dnn} with a higher number of hidden layers and neurons. We notice that the \gls{ber} decreases as the size of the bottleneck and the depth of the head model increases. This performance enhancement, however, comes to the detriment of computational load and feedback overhead. 

\begin{table}[!h]
\centering
\small
\begin{tabular}{|c|l|l|l|}
\hline
\textbf{\begin{tabular}[c]{@{}c@{}}BW \\ (MHz) \end{tabular}} & \multicolumn{1}{c|}{\textbf{\begin{tabular}[c]{@{}c@{}}Full\\ Model\end{tabular}}} & \multicolumn{1}{c|}{\textbf{\begin{tabular}[c]{@{}c@{}}$\boldsymbol{|\mathcal{B}|}$\end{tabular}}} & \textbf{BER} \\ \hline
 & \cellcolor[HTML]{C0C0C0}\textbf{224-28}-28-224 & \cellcolor[HTML]{C0C0C0}28 & \cellcolor[HTML]{C0C0C0}0.0342 \\
 & {\color[HTML]{000000} \textbf{224-896-1792}-1792-896-224} & 1792 & 0.0077 \\
\multirow{-3}{*}{20} & \textbf{224-896-896-448}-448-224-224 & 448 & 0.015 \\ \hline 
 & \cellcolor[HTML]{C0C0C0}\textbf{456-57}-57-456 & \cellcolor[HTML]{C0C0C0}57 & \cellcolor[HTML]{C0C0C0}0.0158 \\
 & {\color[HTML]{333333} \textbf{456-1824-3648}-3648-1824-456} & 3648 & 0.0053 \\
 \multirow{-3}{*}{40} & \textbf{456-1824-1824-912}-912-456-456 & 912 & 0.0083 \\ \hline
 & \cellcolor[HTML]{C0C0C0}\textbf{968-121}-121-968 & \cellcolor[HTML]{C0C0C0}121 & \cellcolor[HTML]{C0C0C0}0.0172 \\
 & \textbf{968-3872-7748}-7748-3872-968 & 7748 & 0.017 \\
 \multirow{-3}{*}{80} & \textbf{968-3872-3872-1936}-1936-968-968 & 1936 & 0.0202 \\
\hline
\end{tabular}
\caption{Impact of the bottleneck placement and size ($\boldsymbol{|\mathcal{B}|}$) on BER, size of \gls{bf} and STAs' computational load. The head model parameters are boldfaced. The highlighted rows are the 3-layer \NW with $K = \sfrac{1}{8}$.\vspace{-0.2cm}}
\label{tab:bottleneck}
\end{table}

On the other hand, \textit{increasing the model parameters does not guarantee to improve the accuracy of the predictions}. For example, for $2 \times 2$ at 20 MHz, the head model with multiply-accumulate (MAC) of 6,641,152 results in a BER of 0.0182 while a model with MAC of 1,612,800 (75\% less computational load) has a lower BER of 0.015. This is due to the model severely overfitting the training data. The results demonstrate that heuristic algorithm simplifies the search process while maintaining acceptable performance, within about $10^{-3}$ of existing approaches.

\begin{table}[!h]
\normalsize
\centering 
\begin{tabular}{|c|c|c|c|c|}
\hline
 & \multicolumn{4}{c|}{\textbf{Signal Bandwidth}}\\ \hline
\textbf{MIMO} & 20 MHz & 40 MHz  & 80 MHz & 160 MHz  \\  \hline
$2 \times 2$ & 0.0202ms & 0.0824ms & 0.3686ms & 1.477ms  \\  \hline
$3 \times 3$ & 0.0459ms & 0.1867ms & 0.8337ms & 3.314ms\\  \hline
$4 \times 4$ & 0.0808ms & 0.3298ms & 1.4782ms & 5.883ms\\  \hline
\end{tabular}
\caption{\NW latency vs MIMO dimensions and bandwidth size. \vspace{-0.2cm}}
\label{tab:fpga}
\end{table}


\textbf{Latency Analysis with FPGA Synthesis.}~While it is sufficient to perform MIMO channel sounding once every 100ms, MU-MIMO channel sounding should be performed at least once every 10ms to account for user mobility, according to \cite{gast2013802}  (see page 73). This implies that keeping the end-to-end BM computation within 10ms latency is fundamental. To this end, we have synthesized in \gls{fpga} the neural networks implemented by \NW. We evaluated the latency with up to 160 MHz, which is the maximum as per the 802.11ax standard, and with MIMO dimensionality up to $4\times 4$\footnote{Although 802.11ax supports MU-MIMO transmissions to up to 8  clients simultaneously, to the best of our knowledge, all the 802.11ax APs currently on the market support only a maximum of 4 spatial streams.}. We considered $K = \sfrac{1}{4}$ compression rate, which results in the lowest BER value as shown in Figure 9.

As target device, we chose a Zynq UltraScale+ XCZU9EG-2FFVB1156, a commonly used System-on-Chip for software-defined radios that is also supported by the OpenWiFi project as part of the ZCU102 evaluation board \cite{openwifigithub}. We chose 5ns as clock period (200 MHz clock frequency), which is the operating clock of the AD9361 transceiver \cite{ad9361} used in OpenWiFi.  We have used a customized library based on high-level synthesis (HLS) designed by us to synthesize the neural networks. HLS allows the conversion of a C++-level description of the \gls{dnn} directly into hardware description language (HDL) code such as Verilog. Therefore, improved latency results could be achieved with more advanced synthesis strategies. Moreover,  better latency results could be achieved by utilizing application-specific integrated circuits (ASICs), which however allow little room for reconfigurability. 

Table \ref{tab:fpga} shows the results obtained through our FPGA synthesis process described above. We notice that by doubling the bandwidth, the latency of the design increases by about 4 times on the average, which is also true when MIMO dimensionality is increased from $2 \times 2$ to $4 \times 4$. In the worst case of 160 MHz and $4 \times 4$ dimensionality, the obtained end-to-end latency is well below the desired 10ms threshold\footnote{The sounding procedure in 802.11ax lasts about 500us \cite{802.11ax}, which makes the overall end-to-end reporting delay below 10ms in the worst case.}.





\section{Concluding Remarks}

We have proposed \NW, a framework to simultaneously reduce the computational load and airtime overhead in modern Wi-Fi networks. We have proposed a new data-driven framework that trains a task-specific \gls{dnn} to output \gls{bf} given the \gls{csi} matrix as input. The key advantage of the \NW is utilizing split \gls{dnn} to insert a bottleneck layer -- which is significantly smaller than the original \gls{csi} -- that (i) enables transferring the computational load of the \gls{sta} to the \gls{ap} side (where computational power is abundant);   (ii) generating a compressed representation of the \gls{bf}, which reduces the feedback airtime. We formulate and solve a bottleneck optimization problem (BOP) to keep computation, airtime overhead and \gls{ber} below application requirements. We have performed extensive experimental \gls{csi} collection in two distinct propagation environments with different bandwidths and number of antennas, and compared the performance with a DNN-based approach and the traditional 802.11 algorithm for BF. Our results have shown that \NW is very effective in reducing the beamforming feedback size and computational complexity by up to 81\%, 84\% with respect to 802.11 while maintaining similar BER  values. For the first time, we have demonstrated that neural networks can be successfully utilized to approximate complex \gls{dsp} operations and thus find the right trade-off between application-specific  requirements and computational/airtime overhead. We believe our  findings could be applied to approximate \gls{dsp} computation beyond Wi-Fi and \gls{bf} compression. We hope that \NW will prompt a new line of research where application-aware neural networks will address network- and device- specific needs more effectively.

\section*{Acknowledgement}

This material is based upon work supported in part by the National Science Foundation (NSF) under Grant No. CNS-2134973, CNS-2134567, CNS-2120447, ECCS-2146754, OAC-2201536, CCF-2218845, and ECCS-2229472, as well as by the Air Force Office of Scientific Research under contract number FA9550-23-1-0261 and by the Office of Naval Research under award number N00014-23-1-2221. The views and opinions are those of the authors and do not necessarily reflect those of the funding institutions or the US Government.


\bibliographystyle{ieeetr}
\bibliography{bib}

\end{document}